\documentclass[runningheads]{llncs}
\usepackage[T1]{fontenc}
\usepackage{graphicx}
\usepackage{multirow}
\usepackage{subcaption}
\usepackage{amsmath}
\usepackage{algorithm}
\usepackage{amssymb}
\usepackage{booktabs}
\usepackage{orcidlink}
\usepackage[backend=biber, style=lncs]{biblatex}
\usepackage{hyperref}
\usepackage{cleveref}
\usepackage{algpseudocode}
\algrenewcommand\algorithmiccomment[1]{//~#1}
\newcommand*\LineComment[1]{\State //~#1}
\newcommand*\ForIn[2]{\ForAll{#1 $\in$ #2}}
\newcommand*\Let[2]{\State #1 $\gets$ #2}
\usepackage{geometry}
\begin{document}
\title{RTS-ABAC: Real-Time Server-Aided Attribute-Based Authorization \& Access Control for Substation Automation Systems}
\titlerunning{RTS-ABAC: Real-Time Server-Aided ABAC for SAS}
\author{%
    Moritz Gstür\orcidlink{0009-0002-4036-9032} \and
    Gustav Keppler\orcidlink{0000-0002-2323-0533} \and
    Mohammed Ramadan\orcidlink{0000-0003-3105-3079} \and
    Ghada Elbez\orcidlink{0000-0003-1137-1782} \and
    Veit Hagenmeyer\orcidlink{0000-0002-3572-9083}
}
\authorrunning{Gstür et al.}
\institute{Karlsruhe Institute of Technology (KIT)
\email{\{moritz.gstuer,gustav.keppler,mohammed.ramadan,ghada.elbez,veit.hagenmeyer\}@kit.edu}}
\maketitle              
\begin{abstract}
    Critical energy infrastructures increasingly rely on information and communication technology for monitoring and control, which leads to new challenges with regard to cybersecurity.
    Recent advancements in this domain, including attribute-based access control (ABAC), have not been sufficiently addressed by established standards such as IEC 61850 and IEC 62351.
    To address this issue, we propose a novel real-time server-aided attribute-based authorization and access control for time-critical applications called RTS-ABAC.
    We tailor RTS-ABAC to the strict timing constraints inherent to the protocols employed in substation automation systems (SAS).
    We extend the concept of conventional ABAC by introducing real-time attributes and time-dependent policy evaluation and enforcement.
    To safeguard the authenticity, integrity, and non-repudiation of SAS communication and protect an SAS against domain-typical adversarial attacks, RTS-ABAC employs mandatory authentication, authorization, and access control for any type of SAS communication using a bump-in-the-wire (BITW) approach.
    To evaluate RTS-ABAC, we conduct a testbed-based performance analysis and a laboratory-based demonstration of applicability.
    We demonstrate the applicability using intelligent electronic devices, merging units, and I/O boxes communicating via the GOOSE and SV protocol.
    The results show that RTS-ABAC is able to secure low-latency communication between SAS devices, as up to 99.82 \% of exchanged packets achieve a round-trip time below 6 ms.
    Moreover, the results of the evaluation indicate that RTS-ABAC is a viable solution to enhance the cybersecurity not only in a newly constructed SAS but also via retrofitting of existing substations.

\end{abstract}
\section{Introduction}
Modern operational technology (OT) increasingly relies on information and communication technology (ICT) for monitoring and control~\cite{Stouffer2023}.
This development leads to new possibilities including the integration of distributed OT into supervisory control and data acquisition (SCADA) systems.
Nevertheless, new challenges arise from the increased usage of ICT in OT systems.
Although a variety of cybersecurity solutions exist for IT, migration of existing approaches to the OT domain may not be a viable solution due to the differing system characteristics, risks, and priorities.
While the prevention of unauthorized access represents the core objective of IT security approaches, OT systems and especially OT-based critical infrastructure prioritize system availability and reliability.

In the energy-related sector, the infrastructure currently transforms from traditional top-down energy transmission and distribution systems to so-called smart grids with bidirectional data and energy flows~\cite{Fang2012}.
The distribution of formerly centralized entities, such as power plants and control centers, necessitates not only changes in energy infrastructure but also leads to an increased reliance on communication solutions.
Historical evidence indicates that economically or politically motivated adversaries pose a risk to OT systems, including energy-related systems \cite{canada2021}.
The \citeauthor{canada2021} \cite{canada2021} published a list of 28 OT-related cybersecurity incidents between 2010 and 2020, including incidents in energy-related sectors.
These incidents comprise 13 state-sponsored incidents, 13 cybercrime incidents, and two incidents perpetrated by thrill-seeking individuals.
The state-sponsored incidents include the Stuxnet malware deployed in Iranian nuclear power and enrichment facilities in 2010 \cite{bbc2010}, the Shamoon malware used against Saudi Aramco in 2012 \cite{reuters2012}, the Blackenergy malware used to attack Ukrainian power distribution systems in 2015 \cite{cisa2021a}, the Industroyer/CrashOverride malware used to shut down remote terminal units of a Ukrainian power transmission facility in 2016 \cite{reuters2016,cisa2021b}, and the Triton/Trisis malware used to attack Triconex safety instrumented system controllers in 2017 \cite{johnson2017}.

In the following, we focus on a particular part of smart electricity grids called substation automation systems (SAS).
An SAS represents the entirety of communication and control equipment of a substation~\cite{Padilla2015}, which is a facility of a high-voltage electricity grid connecting power transmission and distribution lines that use different voltage levels~\cite{oshaSubstation}.
The tasks of an SAS are time-critical and have to be executed reliably, as the electricity sector and its substations are critical infrastructures.
Although standards regarding the communication within an SAS are widely accepted and utilized, cybersecurity continues to confront unresolved challenges.
The IEC 61850 series provides standards for the communication networks of digital energy systems, which ensure seamless communication and interoperability in smart grids~\cite{IEC61850P5}.
However, cybersecurity is not an objective of these standards.
To overcome this problem, the IEC 62351 standard series provides standardized security means for communication compliant to IEC 61850 and a role-based access control (RBAC) model for power systems~\cite{IEC62351P6,IEC62351P8}.

The present paper focuses on enhancing the cybersecurity of SAS communication without compromising its time criticality, as the strict time constraints of the low latency communication in an SAS are key factors for the cybersecurity \cite{Ishchenko2018,Elbez2019}.
Thus, we propose a real-time server-aided attribute-based authorization and access control (RTS-ABAC) approach.
The objective of RTS-ABAC is to provide secure protocols and algorithms for SAS communication, which satisfy the SAS security requirements, including integrity, authenticity, access control, and non-repudiation.
Furthermore, RTS-ABAC takes the specific characteristics, risks, and priorities of OT, ICS, and SAS into account.
To address the aforementioned objectives and considerations, the present paper comprises the following main contributions:
\begin{description}
    \item[C1] A novel function-oriented component-based SAS communication architecture, which integrates SAS components as well as components responsible for authentication, authorization, and access control.

    \item[C2] Two novel strategies for the employment of ABAC in an SAS:
    \begin{description}
        \item[C2.1] A strategy for the representation, specification, and classification of ABAC policies, which introduces the notion of time-dependent attributes and policies.
        \item[C2.2] A strategy for the evaluation of time-dependent access control policies and the enforcement of the respective access control decisions in distributed real-time systems.
    \end{description}

    \item[C3] Two novel communication protocols specifying the message exchanges of RTS-ABAC:
    \begin{description}
        \item[C3.1] A delegated attribute-based authorization protocol responsible for the access control policy creation, management, storage, and distribution.
        \item[C3.2] A delegated decision enforcement protocol responsible for the exchange, verification, and enforcement of access decisions.
    \end{description}

    \item[C4] An experimental evaluation analyzing the performance and applicability of RTS-ABAC:
    \begin{description}
        \item[C4.1] A performance analysis of RTS-ABAC using different authentication algorithms in a testbed based on off-the-shelf hardware.
        \item[C4.2] A laboratory-based demonstration of applicability with industrial SAS devices communicating via the GOOSE and SV protocol.
    \end{description}
\end{description}

The remainder of the paper is organized as follows.
In \autoref{sec:preliminaries}, we present the fundamentals and related literature of RTS-ABAC.
In \autoref{sec:approach}, the architecture and communication protocols of RTS-ABAC are introduced and our novel access control policy specification, classification, and evaluation strategies are discussed.
We present the evaluation of RTS-ABAC, which consists of a performance analysis and a laboratory-based demonstration of applicability, in \autoref{sec:evaluation}.
In \autoref{sec:conclusion}, we present a summary and provide insights into prospective future research.

\section{Preliminaries}
\label{sec:preliminaries}

\subsection{Attribute-Based Access Control (ABAC)}
\label{sec:preliminaries_abac}
ABAC is an access control model enabling access decisions based on attributes associated with subjects, objects, actions, and the environment of a system \cite{JTF2020}.
Within the context of ABAC, an attribute is a characteristic containing information in the form of a name-value pair \cite{Hu2014}.
A policy represents a rule based on which an access decision is taken for specific attributes.
As a consequence, a policy can be seen as a relationship between subject, object, environment, and operation attributes describing under which circumstances the access control mechanism grants or denies an access request.

As the eighth part of the IEC 62351 series provides a RBAC concept for power systems~\cite{IEC62351P8}, RBAC represents the predominant access control model employed in the smart grid domain and related literature.
Since a role can be seen as a subject attribute evaluated by the access control mechanism to take an access decision, RBAC represents a special and limited case of ABAC regarding the attributes used~\cite{Hu2014}.
Accordingly, an advantage of ABAC is the higher flexibility regarding multifactor policy expression.
Moreover, ABAC can take access control decisions based on ad-hoc knowledge.

\subsection{Substation Automation Systems (SAS): Architecture \& Communication Protocols}
\label{sec:preliminaries_sas}
\begin{figure}
    \centering
    \includegraphics[width=0.8\linewidth]{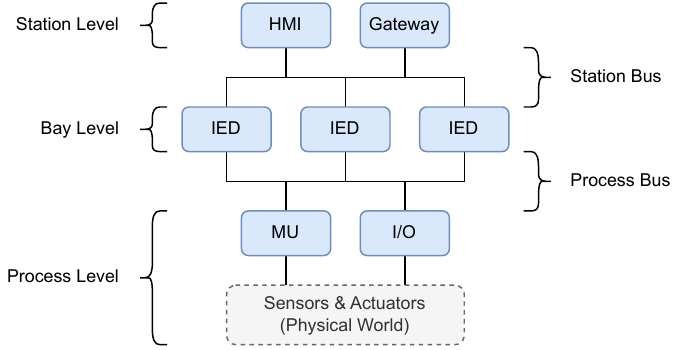}
    \caption{Three-layered architecture of an SAS.}
    \label{fig:substation_architecture}
\end{figure}

RTS-ABAC is tailored to the communication and control systems of digital substations in smart electricity grids.
However, the main concepts may also be applied to other ICSs with similar requirements and constraints.
The SAS architecture shown in \autoref{fig:substation_architecture} is based on the IEC 61850 standards \cite{IEC61850P5} and consists of three layers called station, bay, and process level.
Each of the layers consists of different devices and provides different control and automation functions.
Devices restricted to the transformation and provision of measurement and control values are referred to as merging units (MU).
Intelligent electronic devices (IED) at bay level supervise the operation of lower-level devices.
Human machine interfaces (HMI) as well as wide area network (WAN) gateways are required for on-site and remote monitoring and control of an SAS.

The types of messages exchanged between devices of an SAS are shown in \autoref{tab:message_types}.
In addition to the IEC 61850-5 message types and performance classes \cite{IEC61850P5}, the table shows the associated SAS protocols of each message type \cite{IEC61850P8}.
Communication within a substation is segregated into two networks: the process bus and the station bus.
The process bus operates using protocols based on Ethernet multicast or broadcast communication.
The generic object oriented substation event (GOOSE) protocol is used to transmit commands and statuses like protection trips and interlocking signals.
The sampled values (SV) protocol is used by MUs to transmit digitized measurements, such as current and voltage measurements, to IEDs at bay level.
The real-time nature of these protocols necessitates time synchronization, provided by the (simple) network time protocol (NTP / SNTP) and the precision time protocol (PTP).
The communication at the station bus is session-based unicast communication with less strict time requirements compared to the process bus.
The manufacturing message specification (MMS) protocol is used for station bus communication in an SAS.
The configuration of data exchanges, for all buses and protocols, is formally defined by the system configuration description language (SCL), specified in IEC 61850-6~\cite{IEC61850P6}.
\begin{table}
    \centering
    \small
    \caption{Types of messages exchanged between SAS devices.}
    \label{tab:message_types}
    \begin{tabular}{c c c c c}
    \toprule

    Message & Message & Performance & Transfer & SAS\\
    Name & Type & Class & Time & Protocol\\
    \toprule
    \multirow{3}{*}{Fast Messages} & \multirow{2}{*}{1A (Trip)} & P1 & $\leq 3~ms$ & \multirow{3}{*}{GOOSE}\\
                                  &                            & P2 & $\leq 10~ms$ &\\
    \cmidrule(lr){2-4}
                                  & 1B (Others)                & P3 & $\leq 20~ms$ &\\
    \midrule
    Medium Speed Messages & 2 & P4 & $\leq 100~ms$ & MMS\\
    \midrule
    \multirow{2}{*}{Low Speed Messages} & \multirow{2}{*}{3} & P5 & $\leq 500~ms$ & \multirow{2}{*}{MMS}\\
                                        &                    & P6 & $\leq 1 000~ms$ &\\
    \midrule
    \multirow{2}{*}{Raw Data Messages} & \multirow{2}{*}{4} & P7 & $\leq 3~ms$ & \multirow{2}{*}{SV}\\
                                       &                    & P8 & $\leq 10~ms$ &\\
    \midrule
    File Transfer Functions & 5 & P9 & $\leq 10 000~ms$ & MMS\\
    \midrule
    Command Messages \& & \multirow{3}{*}{6} & P10 & $\leq 500~ms$    & \multirow{3}{*}{i.a. PTP \& SNTP}\\
    File Transfer       &                    & P11 & $\leq 1 000~ms$  &\\
    with Access Control &                    & P12 & $\leq 10 000~ms$ &\\
    \bottomrule
    \end{tabular}
\end{table}

\subsection{Related Work}
\label{sec:preliminaries_related_work}
An authenticated communication approach for network packets between IEDs and MUs is presented by \citeauthor{Ishchenko2018} \cite{Ishchenko2018}.
Based on the concept of retrofitting, a bump-in-the-wire (BITW) device called security filter is presented as an add-on device between IEDs and Ethernet-based communication busses using the GOOSE or SV protocol.
The security filter appends MAC tags to outgoing messages of an IEDs and verifies incoming MAC tags.
Consequently, the communication busses are secured against unauthenticated messages achieving the security goals integrity and authenticity.
The authors show that the security filter is able to meet the IEC 61850 performance requirements of GOOSE and SV \cite{IEC61850P5} using a hash message authentication code (HMAC) and galois message authentication code (GMAC) algorithm even on commodity off-the-shelf ARM hardware.
The architecture and security procedures of RTS-ABAC are inspired by the security filter \cite{Ishchenko2018}.
We use the concept of authenticated communication as a foundation for secure communication in substations.
However, RTS-ABAC is as an authentication-dependent but scheme-agnostic approach that safeguards not only authentication-related but also authorization-related security requirements.
Moreover, RTS-ABAC extends the employed access control from identity-based to attribute-based authorization to enable more complex access control policies.

An authentication and encryption approach for GOOSE and SV communication is presented by \citeauthor{Rodriguez2021} \cite{Rodriguez2021}.
The authors propose an FPGA-implementable hardware architecture for the encryption and authentication of GOOSE and SV packets at wire-speed conforming to IEC 62351:2020~\cite{IEC62351P6}.
The authors conduct an evaluation of the presented architecture using simulation-based and hardware-based timing results.
As stated by the authors, the hardware implementation is able to process GOOSE and SV packets with a fixed latency in the order of microseconds.
Besides securing the intra-substation communication based on the GOOSE and SV protocol, RTS-ABAC extends the idea of providing integrity, authenticity, and non-repudiation to any type of SAS communication.
In contrast to the hardware architecture by \citeauthor{Rodriguez2021} \cite{Rodriguez2021}, RTS-ABAC is software-based.
Our software-based solution achieves flexibility and interoperability with regard to different ICS environments including different protocols and algorithms used.
Moreover, RTS-ABAC is cost-efficient and requires neither special equipment nor specialized human operators due to its deployment onto inexpensive off-the-shelf hardware. 

An access control approach driven by ABAC policies for smart grid systems including substations is presented by \citeauthor{Ruland2018} \cite{Ruland2018}.
The presented access control approach is based on an access control firewall, which splits the station bus and connects the outer and inner station bus by processing access requests of connected devices.
The access control firewall enforces access request decisions based on ABAC policies.
The policies used in the presented approach are defined, communicated, and evaluated using the extensible access control markup language (XACML) \cite{Oasis2013}.
Accordingly, the access request decisions are made by a policy decision point (PDP) that can either be part of the access control firewall or be implemented as an external server on the outer station bus.
RTS-ABAC employs ABAC similarly to the access control approach presented by \citeauthor{Ruland2018} \cite{Ruland2018}.
In addition to the employment of ABAC to secure the communication between devices on the station bus, RTS-ABAC controls access requests to any device within the substation that requires access control.
For this purpose, we use a distributed ABAC firewall instead of a single firewall.
As a consequence, RTS-ABAC does not represent a communication bottleneck or single point of failure of an SAS.

A real-time capable ABAC approach is presented by \citeauthor{Burmester2013} \cite{Burmester2013}.
According to the authors, employing ABAC in real-time availability scenarios can be challenging due to the dynamic and large event space determining the attribute values.
In these real-time availability scenarios, events threatening the system state might not be addressed within strict time limits if attribute values are not available in time.
For this purpose, the authors propose an extended ABAC model that is based on time-dependent attributes to support availability within the strict time constraints of cyber-physical systems.
The authors demonstrate the real-time ABAC for IP multicast in Trusted Computing (TC) compliant networks.
Moreover, as stated by the authors, the extended ABAC model is applicable to an SAS and medical cyber-physical systems.
The access policy classification and evaluation approach proposed in our paper is inspired by the authors' concept of real-time attributes \cite{Burmester2013}.
While relying on a similar concept of real-time attributes, RTS-ABAC differs in its handling of these attributes.
In our approach, real-time attributes together with static attributes form dynamic and static ABAC policies, i.e., access control policies whose evaluation does or does not rely on time-variable subject, object, environment, or action attributes.
Consequently, we extend the notion of real-time attributes to dynamic and static ABAC policies.
Moreover, we introduce novel strategies to represent, specify, classify, evaluate, and enforce these novel policies.

A distributed RBAC (DRBAC) approach for subscription-based remote network services is presented by \citeauthor{Ma2006} \cite{Ma2006a,Ma2006}.
DRBAC is a distributed authentication and role-based authorization framework.
The distributed authentication is realized by delegating the authentication of users to the corresponding subscribing institutions.
The DRBAC framework extends conventional RBAC by adding the concept of distributed roles shared by the resource provider and resource subscribers.
Thus, it enables access control policies associated with distributed roles rather than subject identities, which leads to an increase in scalability and manageability of access control.
Moreover, the authors state that the DRBAC approach supports temporal, contextual, and cardinality constraints to enhance the semantic expressiveness of access control and enable the definition of higher-level organizational policies.
Our authentication and authorization approach utilizes the concept of delegation presented by \citeauthor{Ma2006} \cite{Ma2006a,Ma2006}.
While the authors illustrate the concept of delegation within the context of subscription-based remote network service environments, RTS-ABAC employs delegation in substations.
RTS-ABAC not only relies on the concept of delegation for authentication but also for authorization and access control via PDPs that make access control decisions for resource requests in place of SAS devices.

An ABAC approach for the data distribution service (DDS), a data-centric machine-to-machine publisher-subscriber standard, is presented by \citeauthor{Kim2022} \cite{Kim2022}.
Due to an increasing utilization of publisher-subscriber communication across different domains, including the Internet of things (IoT) as well as OT systems, DDS security is increasingly relevant.
To tackle this problem, the authors present an approach enhancing the authorization of DDS by employing ABAC.
The authors evaluate the approach by implementing it based on XACML and applying it to a patient monitoring system in the healthcare domain.
The presented implementation is capable of handling 150 access requests simultaneously with an average communication time of 4.95 ms.
According to the authors, the approach is applicable to the healthcare domain and satisfies the transfer time requirements of all message types of IEC 61850-5 \cite{IEC61850P5} except for type 1A and type 4.
While \citeauthor{Kim2022} demonstrate the applicability to medium and low speed IEC 61850-5 message types \cite{Kim2022}, RTS-ABAC provides security to all SAS communication protocols, including the low-latency message types 1A and 4.
Moreover, RTS-ABAC highlights the importance of not only authorization but also authentication in order to enhance the security of low-latency communication in an SAS in a timely and reliable manner.

\section{RTS-ABAC Approach}
\label{sec:approach}
In this section, we introduce our real-time server-aided attribute-based authorization and access control (RTS-ABAC) approach.
RTS-ABAC represents a significant advancement over our initial server-aided attribute-based authorization and access control (SABAAC) approach for SAS \cite{Gstuer2025}.
RTS-ABAC safeguards the principle of least privilege and separation of duties.
Furthermore, as all message exchanges are secured using cryptographic authentication mechanisms, RTS-ABAC safeguards the data frame payload integrity, data frame sender authenticity, and data frame authorship non-repudiation.
However, the availability of secure cryptographic authentication mechanisms is assumed by RTS-ABAC, as it is beyond the scope of the present paper.
To satisfy the strict time constraints of the SAS domain, the expressive and flexible yet computationally expensive ABAC policies are handled in a server-aided manner.

\begin{figure}
    \centering
    \includegraphics[width=1.0\linewidth]{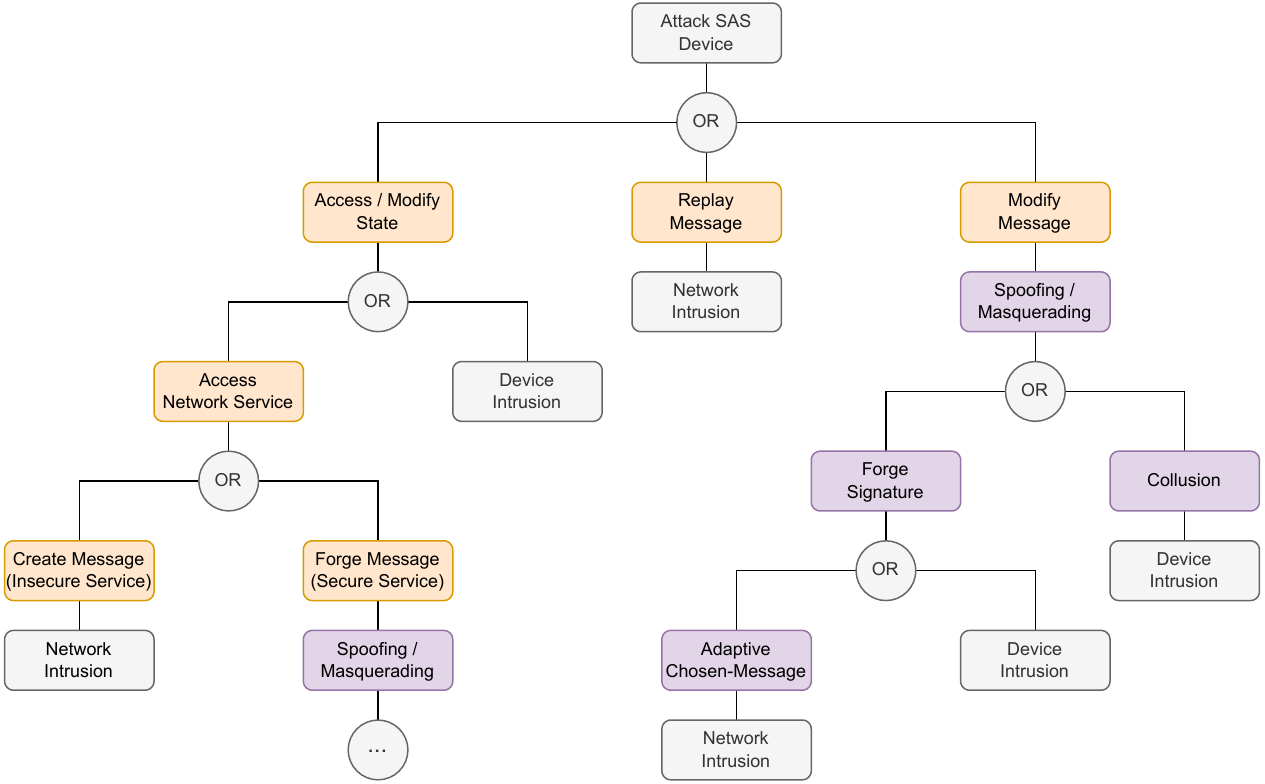}
    \caption{Attack tree comprising cyberattacks that endanger SAS devices and message exchange protocols.}
    \label{fig:attack_tree}
\end{figure}

As the objective of an adversary is typically unachievable by a single operation but rather requires a sequence of attacks to be achieved, we model the objectives of RTS-ABAC as an attack tree, which is shown in \autoref{fig:attack_tree}.
The attacks depicted in orange are integrity-focused, i.e., aim to disturb the system's integrity by transitioning the system into a state that is beneficial for an adversary.
The attacks depicted in purple are authenticity-focused and enable an adversary to impersonate a legitimate subject of the system, thereby abusing the subject's granted privileges.
We identify two primary types of adversaries for RTS-ABAC.
The first adversary has the objective of compromising the system by disrupting communication between two or more legitimate system subjects.
In order to compromise the communication integrity, the adversary may either replay messages that have been captured on the network, or modify them.
While replaying only requires the adversary to be able to eavesdrop on communication, message modification additionally requires the adversary to masquerade as a legitimate subject.
This can be achieved by breaking authenticity protection mechanisms, such as digital signatures, or by colluding with other adversaries, such as infiltrated system devices.
The second adversary has the objective of compromising the system by directly attacking the SAS devices.
This adversary may seek to either disrupt the device's availability or compromise its integrity.
A device's integrity can be compromised by modifying its state using accessible service interfaces.
Accordingly, an adversary may either create a legitimate request, in case of an unsecured service, or forge a request.

\subsection{Authorization \& Access Control Architecture}
\label{sec:approach:architecture}
\begin{figure}
    \centering
    \includegraphics[width=1.0\linewidth]{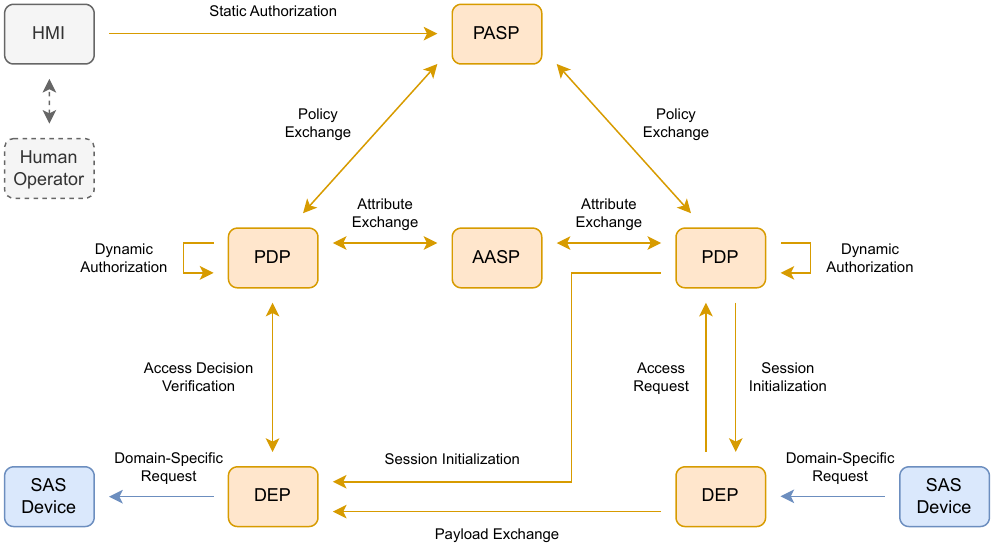}
    \caption{Function-oriented component-based architecture of the RTS-ABAC approach.}
    \label{fig:protocols_overview}
\end{figure}

RTS-ABAC is based on a function-oriented component-based architecture.
An overview of the architecture, components, and protocols is shown in \autoref{fig:protocols_overview}.
The component-based architecture consists of four functional units.
The functional units are inspired by the access control mechanism functional points presented by \citeauthor{Hu2014} \cite{Hu2014}.
Each functional unit is represented by a component that offers a set of services.
The four components of the architecture are defined below:
\begin{description}
    \item[Policy Administration \& Storage Point (PASP)] The PASP offers services for policy creation, storage, management, and distribution.
    Accordingly, it provides interfaces for policy management services to human operators.
    Moreover, the PASP acts as a repository for created policies and changes to policies.

    \item[Policy Decision Point (PDP)] The PDP takes access control decisions by evaluating policies.
    The PDP takes an access decision on request of another device and provides the decision to the requestor.
    Thus, it is responsible for the policy and attribute evaluation workflow, including the retrieval of required attributes and speedup techniques such as access decision caching and policy evaluation precomputation.
    Moreover, the PDP incorporates the services provided by the context handler, which is introduced by \citeauthor{Hu2014} \cite{Hu2014}.

    \item[Attribute Administration \& Storage Point (AASP)] The AASP offers services for attribute creation, management, and distribution.
    The AASP supports the authorization procedure by providing attributes to the PDPs.
    Accordingly, the AASP corresponds to the Policy Information Point (PIP) introduced by \citeauthor{Hu2014} \cite{Hu2014}.

    \item[Decision Enforcement Point (DEP)] The DEP enforces access control decisions by controlling access to protected objects.
    In contrast to the PEP introduced by \citeauthor{Hu2014} \cite{Hu2014}, the DEP neither has direct access to policies nor attributes but enforces policies indirectly via access decisions taken by a PDP.
    In addition to the tasks of conventional PEPs, our DEPs are responsible for providing authentication-related services to SAS devices.
    The services for SAS devices are provided in the form of a BITW solution, i.e., these services are provided to the corresponding service consumers automatically and invisibly.
\end{description}

\subsection{Policy Specification}
\label{sec:approach:specification}
RTS-ABAC features a novel strategy for the representation and specification of access control policies.
The policies are enforced for data frame flows, i.e., sets of data frames with common properties passing through the SAS network.
Accordingly, the objective of an access control policy is twofold.
On the one hand, a policy has to contain information that specifies if the policy is applicable to a certain data frame flow.
On the other hand, a policy must specify which actions should be taken for a matching data frame flow.

\subsubsection{Access Control Policy:} $\rho = (action_{\rho}, flow_{\rho}, auxiliary_{\rho})$.\\
A policy $\rho$ is represented by a three-tuple, which contains the $action$ to be taken, the $flow$-specifying pattern, and a set of attribute-based $auxiliary$ predicates specifying additional non-$flow$-related system characteristics.

\subsubsection{Action:} $action \in \{GRANT, DENY\}$.\\
The action of a policy specifies whether a matching data frame flow should be granted or denied.
A data frame from a granted flow may be delivered to an SAS device, whereas a data frame from a denied flow is dropped by a DEP.
The default action is $DENY$, which leads to dropping of all data frames of non-explicitly granted flows.

\subsubsection{Flow Pattern:} $flow = \{p_1, \dots, p_n,\}$, $p_i(f) \in \{TRUE, FALSE\}$.\\
The flow pattern of a policy specifies whether the policy is applicable to a specific data frame or not.
It consists of a set of frame predicates $p_1, \dots, p_n$.
A frame predicate $p_i$ is a function that assigns a boolean value to an arbitrary data frame.
A predicate $p_i$ matches a frame $f$, if $p_i(f) = TRUE$.
A policy is applicable to a specific data frame if all frame predicates match the frame.
In other words, a policy $\rho = (action_{\rho}, flow_{\rho}, auxiliary_{\rho})$ is applicable to a frame $f$ if the following equation holds:
\[
    \forall p_i \in flow_{\rho}: p_i(f) = TRUE \iff flow_{\rho}(f) = p_1(f) \land \dots \land p_n(f) = TRUE
\]

\begin{figure}
    \centering
    \includegraphics[width=0.9\linewidth]{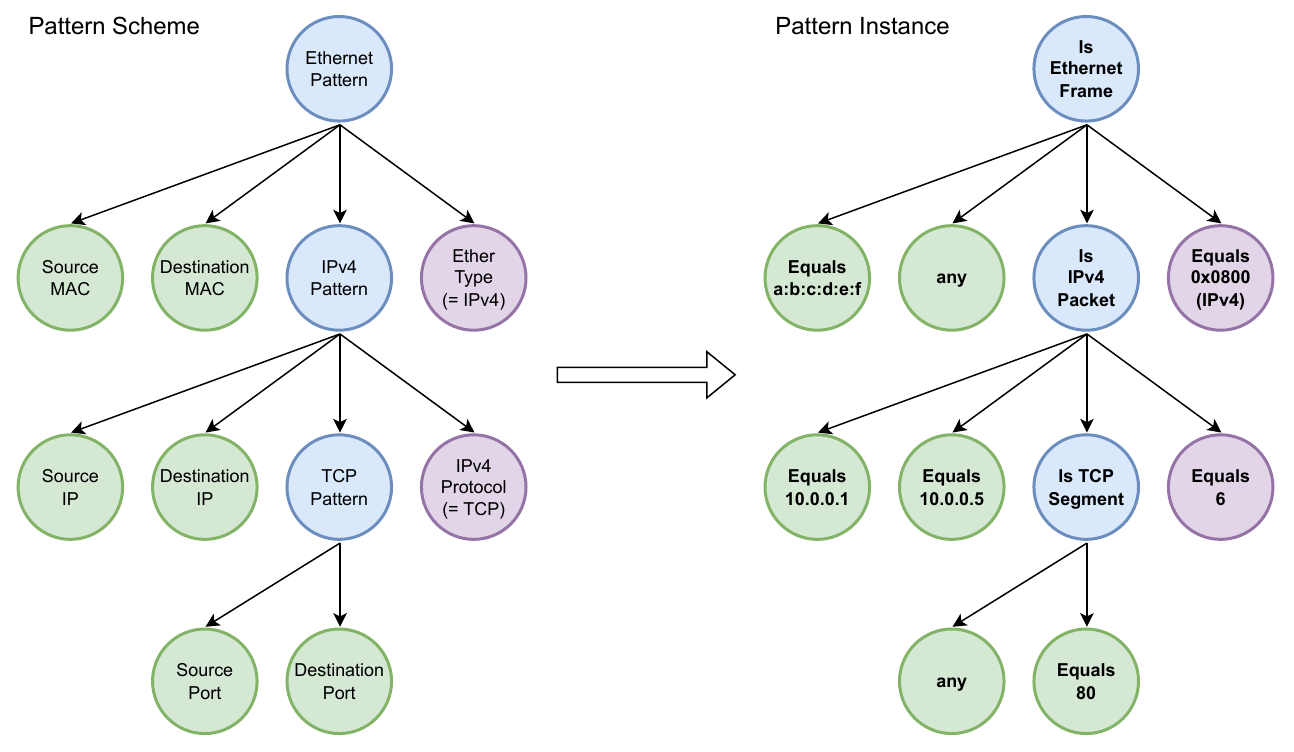}
    \caption{Scheme and instance of a tree-like flow pattern representing the flow predicates of a TCP segment.}
    \label{fig:flow_pattern_scheme_instance}
\end{figure}

To increase the efficiency of the predicate evaluation and simplify the flow specification, each flow pattern is represented as a directed graph.
By enclosing one pattern into another pattern, tree-like flow patterns can be constructed.
A tree-like flow pattern of a transmission control protocol (TCP) segment is shown in \autoref{fig:flow_pattern_scheme_instance}.
Each node of the tree-like pattern represents one flow predicate.
RTS-ABAC distinguishes between three types of flow predicates.
The predicates depicted in blue are hierarchy predicates.
Hierarchy predicates determine the structure of the tree, i.e., these predicates determine which other predicates can be added to the tree-like pattern.
The predicates depicted in purple are so-called hierarchy-constrained predicates.
Hierarchy-constrained predicates are non-parametric and are derived from the structure of the tree.
The predicates depicted in green are parametric predicates.
Parametric predicates can be configured to match the expected data frame flows within an SAS.

\begin{figure}
    \centering
    \includegraphics[width=0.8\linewidth]{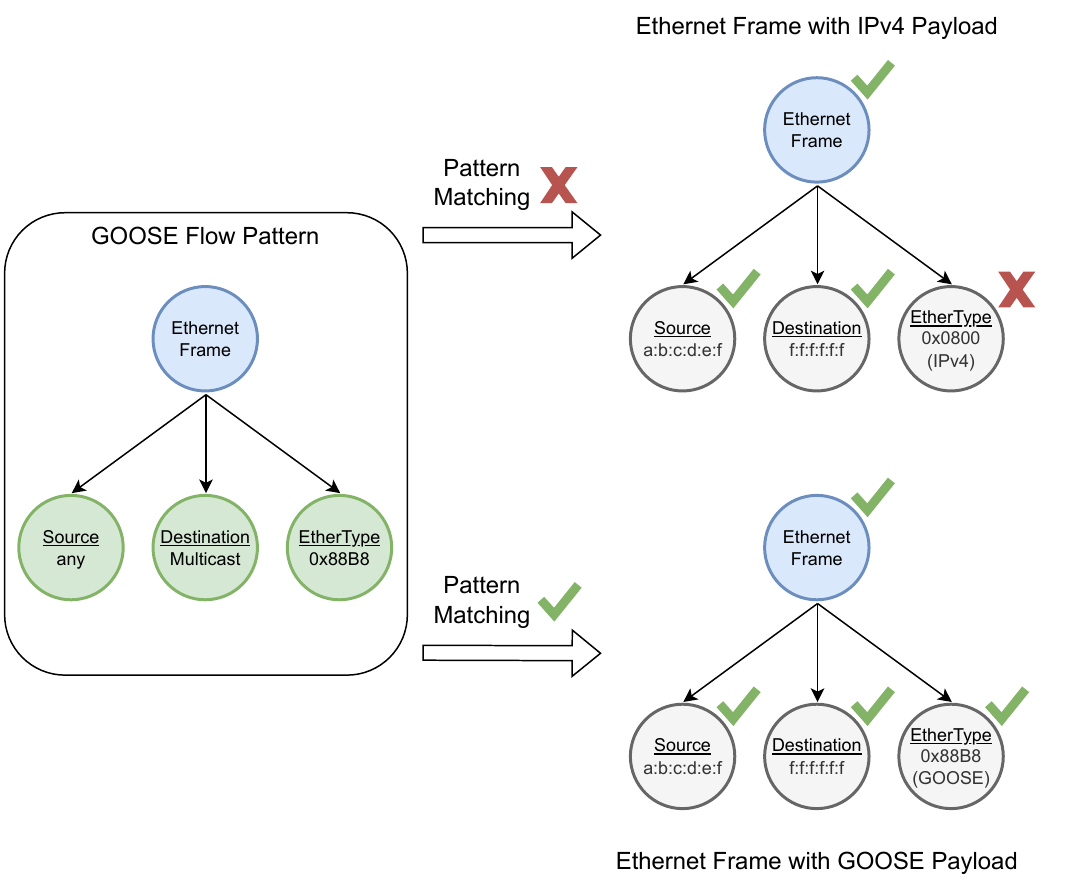}
    \caption{Pattern matching of a GOOSE flow pattern and two Ethernet access request patterns.}
    \label{fig:flow_pattern_matching_full}
\end{figure}

\begin{figure}
    \centering
    \includegraphics[width=0.9\linewidth]{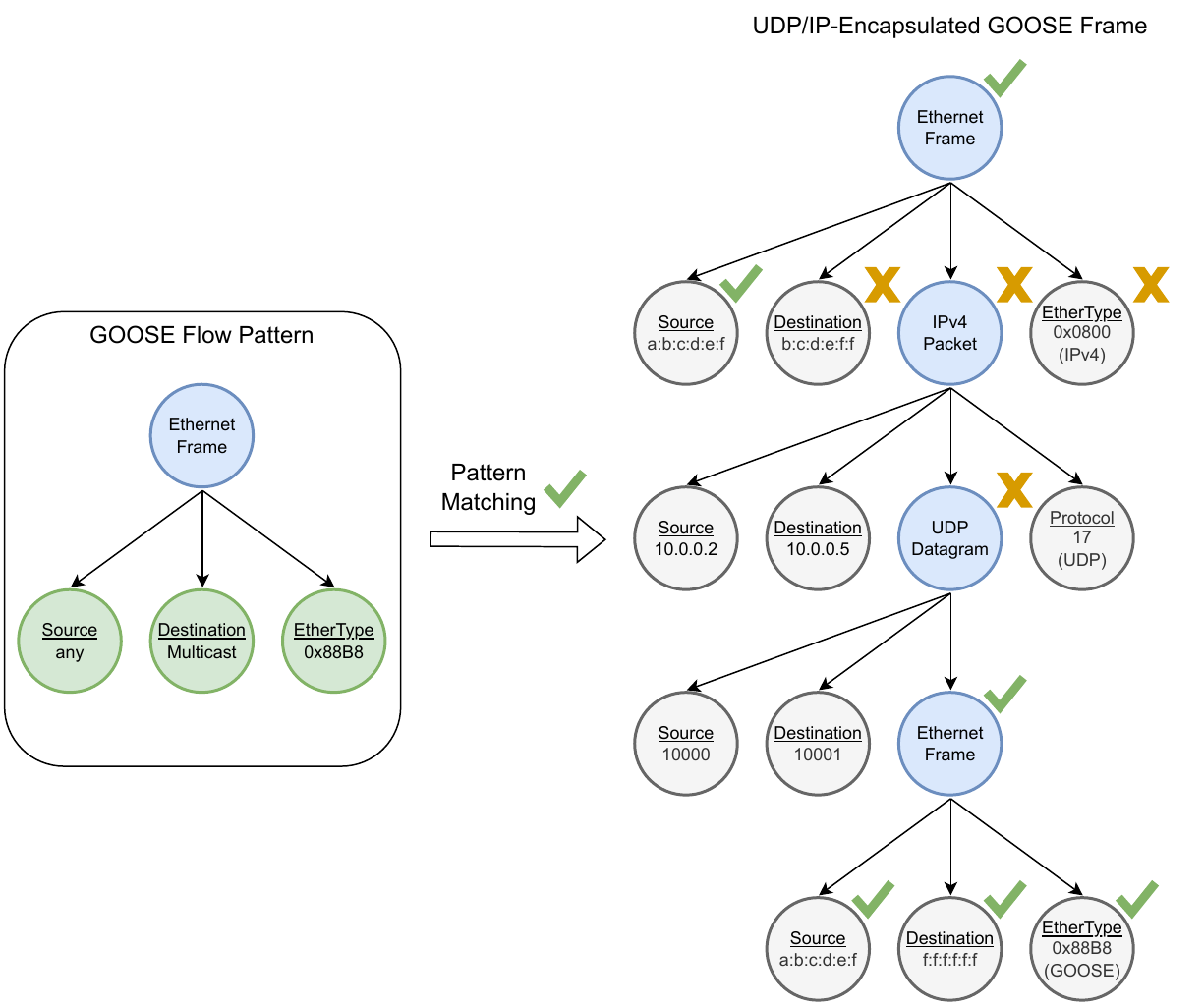}
    \caption{Pattern matching of a GOOSE flow pattern and a UDP/IP-encapsulated GOOSE access request pattern.}
    \label{fig:flow_pattern_matching_nested}
\end{figure}

To check if a flow pattern matches a frame, the frame is transformed into its pattern representation.
The pattern representing a frame is referred to as access request pattern.
While the flow pattern consists of predicates, the access request pattern consists of constant values or facts.
A flow pattern matches a frame's access request pattern, if all predicates of the flow pattern are true for their corresponding constant values of the access request pattern.
To identify the pairs of predicates and facts that belong together, hierarchy predicates are used as anchor points of the patterns.
\autoref{fig:flow_pattern_matching_full} shows a pattern matching process of a GOOSE flow pattern and two access request patterns representing Ethernet frames.
The hierarchy predicates and facts used as anchor points for the pattern matching are depicted in blue.
The remaining facts of the access request patterns are depicted in gray.
As modern network communication is based upon layering of protocols, RTS-ABAC supports matching of nested patterns.
A pattern matching process of a GOOSE flow pattern and a UDP/IP-encapsulated GOOSE access request pattern is shown in \autoref{fig:flow_pattern_matching_nested}.
To find the nested pattern match, RTS-ABAC performs pattern matching of the flow pattern at all anchor points of the access request pattern.
The pattern matching ends if either a match has been found or the flow pattern does not match at any of the anchor points.

\subsubsection{Auxiliary Precondition:} $auxiliary = \{a_1, \dots, a_n\}$, $a_i(system) \in \{TRUE, FALSE\}$.\\
The auxiliary precondition or system precondition of a policy specifies constraints for non-flow-related system attributes that have to be taken into account in order to apply an access control policy.
The auxiliary precondition consists of a set of auxiliary predicates $a_1, \dots, a_n$.
An auxiliary predicate $a_i$ is a function that assigns a boolean value to the system state at a given point in time.
The system state is represented by the set of all non-flow-related system attributes $ATT_{system} \subseteq system$ available at a given time.
Accordingly, an auxiliary predicate is a function
\[
    a_i(system) = a_i(k_1, \dots, k_l \in ATT_{system}) = b \in \{TRUE, FALSE\}\\
\]
that maps an arbitrary number of system attributes $k_1, \dots, k_l$ to a boolean value $b$.
The set of system attributes $ATT_{a_i} := \{k_1, \dots, k_j\} \subseteq ATT_{auxiliary_{\rho}} \subseteq ATT_{system}$ denotes the attributes required for the evaluation of the system predicate $a_i$.
The set $ATT_{auxiliary_{\rho}} := \bigcup_{i=1}^{n} ATT_{a_i}$ denotes all system attributes required for the evaluation of the system precondition $auxiliary_\rho = \{a_1, \dots, a_n\}$ of a policy $\rho$.
The auxiliary precondition is said to match the system state at a given time if all auxiliary predicates evaluate to true.
A policy $\rho = (action_{\rho}, flow_{\rho}, auxiliary_{\rho})$ is applicable in a system if the auxiliary precondition matches the current system state, i.e., a policy is applicable if the following equation holds:
\begin{align*}
    auxiliary_{\rho}(system) & = auxiliary_{\rho}(ATT_{system}) \\
    &= auxiliary_{\rho}(ATT_{auxiliary_{\rho}}) \\
    &= \bigwedge_{a_i \in auxiliary_{\rho}} a_i(ATT_{auxiliary_{\rho}}) \\
    &= \bigwedge_{a_i \in auxiliary_{\rho}} a_i(ATT_{a_i}) \\
    &= a_1(ATT_{a_1}) \land \dots \land a_n(ATT_{a_n}) = TRUE
\end{align*}

\begin{figure}
    \centering
    \includegraphics[width=0.70\linewidth]{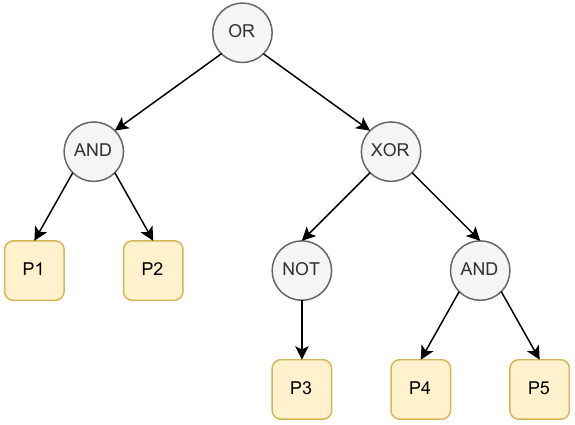}
    \caption{Auxiliary precondition represented by a boolean binary expression tree.}
    \label{fig:boolean_binary_expression_tree}
\end{figure}

While the logical evaluation of numerous predicates without any hierarchy is an efficient task for modern computers, human operators might experience problems with this representation when specifying complex auxiliary preconditions.
To overcome this issue, RTS-ABAC features an auxiliary precondition representation based on boolean binary expression trees.
This representation is referred to as auxiliary predicate tree.
\autoref{fig:boolean_binary_expression_tree} shows a predicate tree of an auxiliary precondition consisting of five predicates.
The inner nodes of a predicate tree are unary or binary logical connectives, including AND, OR, XOR, and NOT.
The leaves of a predicate tree are auxiliary predicates.
The predicate tree can be transformed into the conjunctive form by transforming the tree into a logical expression, calculating the conjunctive normal form (CNF) of this expression, and merging the disjunctions of predicates into new atomic predicates.

\subsection{Policy Classification}
\label{sec:approach:classification}
The dynamic or ad-hoc evaluation of ABAC policies allows the utilization of attributes from a time-variable environment.
A real-time attribute represents an attribute whose value is time-dependent \cite{Burmester2013}.
Given an attribute evaluation function $E_{ATT}$ and a point of time $t$, the value $\lambda_k$ of a real-time attribute $k$ is defined by $\lambda_k := E_{ATT}(k, t)$.

Each policy $\rho$ is related to a set of attributes $ATT_{auxiliary_{\rho}}$.
The auxiliary attributes $ATT_{auxiliary_{\rho}}$ consist of all non-flow-related system attributes required for the evaluation of the auxiliary predicates $a_1, \dots, a_n$.
To handle policies based on their degree of time-variability, we classify policies as follows:
\begin{description}
    \item[Static Policy] A static policy $\rho$ is an ABAC policy whose evaluation does not rely on time-variable attributes.
    A policy $\rho$ is static if $\forall k \in ATT_{auxiliary_{\rho}}: \forall t_i,t_j: E_{ATT}(k,t_i) = E_{ATT}(k,t_j)$.
    Consequently, access decisions can be cached.
    Due to the non-frequent attribute retrieval, non-frequent evaluation, and access decision caching, static policies are a viable solution for low latency message exchange.
    A static policy is also referred to as non-real-time policy.

    \item[Dynamic Policy] A dynamic policy $\rho$ is an ABAC policy whose evaluation relies on at least one time-variable attribute.
    A policy $\rho$ is dynamic if $\exists k \in ATT_{auxiliary_{\rho}}: \exists t_i \neq t_j: E_{ATT}(k,t_i) \neq E_{ATT}(k,t_j)$.
    Due to the time-variable evaluation of dynamic policies, access decisions must have a limited time of validity that corresponds to the change rate of the underlying attribute values.
    As a result, caching of access decisions that are based on dynamic policies should be avoided.
    A dynamic policy is also referred to as real-time policy.
\end{description}

\subsection{Policy Evaluation \& Decision Enforcement}
\label{sec:approach:evaluation}
While the access control policy may depend upon the current system state, the corresponding access control decision is static during a specified period of validity.
As a consequence, the process of handling access control policies can be divided into two related processes:
The PDP-driven dynamic authorization derives an access control decision from an access control policy for the current system state.
The DEP-driven decision enforcement utilizes the result of the dynamic authorization to enforce the policies in a timely manner.

\subsubsection{Access Control Decision:} $decision_{\rho} = \{flow_\rho, action_\rho, nexthop, validity\}$.\\
An access control decision consists of a $flow$-specifying pattern, an $action$ to be taken, a set of $nexthop$ DEPs, and a period of $validity$.
Accordingly, an access decision is valid for a specific data frame $flow$ in a specific system during a specified period of time.
For each access policy $p$ there exists exactly one access decision $decision_{\rho}$ at any point in time.

\subsubsection{Dynamic Authorization:} $E_{POL}: Policy \times System \rightarrow Decision$.\\
The policy evaluation of RTS-ABAC is server-aided, i.e., it is delegated to a PDP.
To derive a decision from a policy at the PDP, the dynamic authorization function $E_{POL}$ is used.
The dynamic authorization function is defined as follows:
\[
    E_{POL}(\rho, s_t) =
    \begin{cases}
        (flow_\rho, action_\rho, nexthop, validity) & \text{, if } \forall a_i \in auxiliary_{\rho}: a_i(s_t) = TRUE,\\
        (flow_\rho, DENY, \emptyset, validity) & \text{, otherwise.}
    \end{cases}
\]
Thus, if the auxiliary predicates $a_1, \dots, a_n$ of the policy $\rho$ match the current system state $s_t$, the policy is applicable to matching data frame flows in the system.
To evaluate the auxiliary predicates, the PDP fetches the auxiliary attributes $ATT_{auxiliary_{\rho}} \subseteq ATT{s_t}$ prior to the evaluation of $a_1, \dots, a_n$.
The validity of an access decision of a policy $\rho$ at a specific time $t$ equals the minimal validity of the attribute values in $\{\lambda_k = E_{ATT}(k, t) | k \in ATT_{auxiliary_{\rho}}\}$.
Consequently, the validity of an access decision of a dynamic policy is determined by the attribute value that expires first.
The validity of an access decision of a static policy may be limited by specifying a non-attribute-related maximum time of validity, to avoid the utilization of outdated access decisions.

\subsubsection{Decision Enforcement:} $E_{DEC}: Decision \times Frame \times Time \rightarrow (Action, Nexthop)$.\\
A DEP uses an access decision $d_{\rho}$ to enforce a policy $\rho$ in the SAS.
For this purpose, each data frame $f$ traversing a DEP is matched using the flow pattern $flow_\rho$ to identify the corresponding access decision.
The process of enforcing a policy based on a priorly taken access decision is referred to as decision enforcement.
At a DEP, the decision enforcement for a decision $d_{\rho} = \{flow_\rho, action_\rho, nexthop, validity\}$ and a frame $f$ at a point in time $t$ is performed using the decision enforcement function $E_{DEC}$:
\[
    E_{DEC}(d_{\rho}, f, t) =
    \begin{cases}
        (action_\rho, nexthop) & \text{, if } \forall p_i \in flow_{\rho}: p_i(f) = TRUE \land d_{\rho}.valid(t),\\
        (DENY, \emptyset) & \text{, otherwise.}
    \end{cases}
\]
Thus, if the flow predicates match a data frame, and the decision is still valid, the decision enforcement function returns the action to be taken and the DEPs the frame has to be forwarded to.
However, two cases have to be distinguished, if $E_{DEC}(d_{\rho}, f, t) = (action_\rho, nexthop)$:
\begin{itemize}
    \item \textbf{Outgoing Data Frame:} An outgoing data frame $f$ is forwarded to each $DEP_i \in nexthop$, if $action_\rho = GRANT$.
    \item \textbf{Incoming Data Frame:} An incoming data frame $f$ is accepted by $DEP_i$, if $action_\rho = GRANT$ and $DEP_i \in nexthop$.
\end{itemize}
If multiple decisions match a data frame $f$, a DEP chooses the decision corresponding to the most specifically matching flow pattern.
When two flow patterns $flow_\gamma$ and $flow_\delta$ of the access control policies $\gamma$ and $\delta$ match a data frame $f$, $flow_\gamma$ is said to match $f$ more specifically than $flow_\delta$ if $flow_\gamma \supsetneq flow_{\delta}$.
In case of $(flow_{\gamma} \not \supseteq flow_{\delta}) \land (flow_{\gamma} \not \subseteq flow_{\delta})$, the matching flow patterns and their corresponding access control policies and decisions are said to be conflicting.
A DEP may use a so-called composite decision $d_C$ to resolve two or more conflicting access control decisions $d_1, \dots, d_n$:
\[
    d_C =
    \begin{cases}
        (\bigcup_{i=1}^{n}flow_i, GRANT, \bigcup_{i=1}^{n}nexthop_i, \min_{i=1}^{n}validity_i) & \text{, if } \forall d_i \in \{d_1, \dots, d_n\}:\\
        & \quad action_i = GRANT,\\
        (\bigcup_{i=1}^{n}flow_i, DENY, \emptyset, \min_{i=1}^{n}validity_i) & \text{, otherwise.}
    \end{cases}
\]

\subsection{Delegated Attribute-Based Authorization Protocol}
\label{sec:approach:protocol_authorization}
The delegated attribute-based authorization protocol is responsible for the access control policy creation, management, storage, and distribution.
The protocol offers reliable services to entities involved in the policy management process, including human operators or intrusion detection and prevention systems.
As access control policies are classified as either static or dynamic, the protocol has to take the time-variability of access control policies and their corresponding access control decisions into account.
The protocol consists of three phases:
\begin{description}
    \item[Static Authorization] The static authorization is responsible for handling create, read, update, and delete (CRUD) requests for access control policies.
    The involved components and message exchanges are visualized in \autoref{fig:protocols_authorization_static}.
    The static authorization is initiated by a human operator or an external system responsible for the management of policies.
    A PASP processes a CRUD request according to the requested action.
    In case of a read request, the PASP provides the requested access control policy to the requestor.
    Create, update, and delete requests are processed by changing the persistent set of access control policies.
    \begin{figure}
        \centering
        \includegraphics[width=1.0\linewidth]{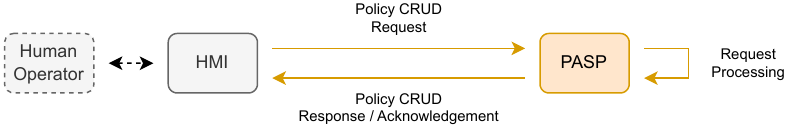}
        \caption{Components and exchanged messages involved in the static authorization process.}
        \label{fig:protocols_authorization_static}
    \end{figure}

    \item[Policy Exchange] After a policy is created, modified, or deleted via static authorization, it has to be shared with the PDPs via policy exchange.
    The policy exchange is an interaction between a PASP and a PDP.
    The interaction is either initiated by a PASP as a result of a static authorization, or on request of a PDP.
    If a static authorization triggers the policy exchange, the PASP sends a policy exchange message to a PDP.
    This type of policy exchange is referred to as incremental policy exchange, i.e., only newly created, modified, or removed policies are exchanged.
    The incremental policy exchange is shown in \autoref{fig:protocols_authorization_policyexchange_incremental}.
    A policy exchange containing all relevant policies can be initiated by a PDP by sending a policy exchange request to a PASP.
    This type of policy exchange is referred to as complete policy exchange.
    The complete policy exchange is shown in \autoref{fig:protocols_authorization_policyexchange_complete}.
    \begin{figure}
        \centering
        \begin{subfigure}[t]{0.48\linewidth}
            \centering
            \includegraphics[width=0.72\linewidth]{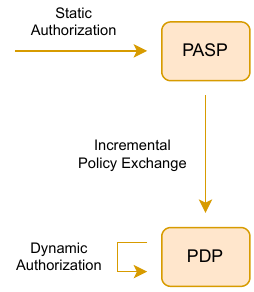} 
            \caption{Incremental policy exchange procedure initiated by a static authorization of a PASP.}
            \label{fig:protocols_authorization_policyexchange_incremental}
        \end{subfigure}
        \hfill
        \begin{subfigure}[t]{0.48\linewidth}
            \centering
            \includegraphics[width=\linewidth]{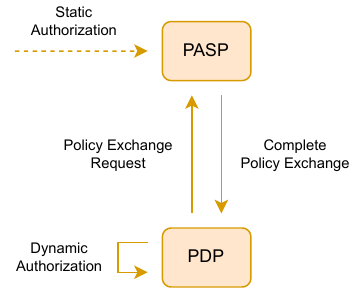}
            \caption{Complete policy exchange procedure initiated by a policy exchange request of a PDP.}
            \label{fig:protocols_authorization_policyexchange_complete}
        \end{subfigure}
        \caption{Components and exchanged messages involved in the policy exchange process.}
        \label{fig:protocols_authorization_policyexchange}
    \end{figure}
    \item[Dynamic Authorization] The dynamic authorization is responsible for deriving an access decision from an access control policy at the PDP.
    The involved components and message exchanges are visualized in \autoref{fig:protocols_authorization_dynamic}.
    \Cref{alg:dynamic_authorization} shows the steps of the dynamic authorization algorithm.
    The dynamic authorization is based on the dynamic authorization function $E_{POL}$.
    Prior to its evaluation, the PDP fetches the required auxiliary attributes from one or multiple AASPs via attribute requests.
    An AASP responds to an attribute request with an attribute resolution message, which contains the values and periods of validity of the auxiliary attributes.
    \begin{algorithm}
        \caption{Dynamic authorization process of a PDP deriving an access control decision for each access control policy in a given system.}
        \label{alg:dynamic_authorization}
        \begin{algorithmic}
        \Function{DynamicAuthorization}{$policies$, $system$}
            \Let{$decisions$}{$\left\{\,\right\}$} \Comment{Initialize an empty set of access decisions}
            \Let{$ATT_{system}$}{$\left\{\,\right\}$} \Comment{Initialize an empty system state}
            \\
            \ForIn{$\rho = (action_{\rho}, flow_{\rho}, auxiliary_{\rho})$}{$policies$}
                \Let{$ATT_{auxiliary_{\rho}}$}{$\left\{\,\right\}$} \Comment{Initialize an empty set of auxiliary attributes}
                \ForIn{$a_i$}{$auxiliary_{\rho}$}
                    \Let{$ATT_{a_i}$}{\Call{RequestAttributes}{$a_i$}} \Comment{Request the attributes from AASPs}
                    \Let{$ATT_{auxiliary_{\rho}}$}{$ATT_{auxiliary_{\rho}} \cup ATT_{a_i}$}
                \EndFor
                \\
                \LineComment{Add the auxiliary attributes of policy $\rho$ to the current system state $ATT_{system}$}
                \Let{$ATT_{system}$}{$ATT_{system} \cup ATT_{auxiliary_{\rho}}$}
                \\
                \LineComment{Derive an access decision from policy $\rho$ for the current system state $ATT_{system}$}
                \Let{$decisions$}{$decisions \cup E_{POL}(\rho, ATT_{system})$}
            \EndFor
            \State \Return $decisions$
        \EndFunction
        \end{algorithmic}
    \end{algorithm}
    \begin{figure}
        \centering
        \includegraphics[width=0.9\linewidth]{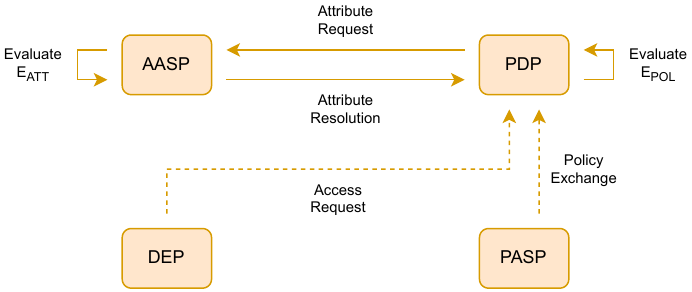}
        \caption{Components and exchanged messages involved in the dynamic authorization process.}
        \label{fig:protocols_authorization_dynamic}
    \end{figure}
\end{description}

\subsection{Delegated Decision Enforcement Protocol}
\label{sec:approach:protocol_enforcement}
The delegated decision enforcement protocol is responsible for the exchange, verification, and enforcement of access decisions.
The workflow of the protocol is divided into three mandatory phases and an optional verification phase.
The phases are defined as follows:
\begin{description}
    \item[Access Request] The access request phase is initiated by a DEP on behalf of an SAS device.
    To identify the corresponding access decision and access policy, the DEP appends the access request pattern of a domain-specific data frame to the access request.

    \item[Session Initialization] The session initialization is performed by a PDP as a response to an access request.
    A session initialization between a requesting DEP and a single $nexthop$ DEP is shown in \autoref{fig:protocols_enforcement_initialization}.
    The PDP identifies all applicable policies via flow pattern matching and fetches the corresponding access decisions from its cache or derives them via dynamic authorization.
    Then the PDP sends session initialization messages to the requesting DEP and the $DEP_i \in nexthop$.
    Each session initialization message contains one or multiple access decisions.
    If no access control policy is applicable, the PDP sends the requesting DEP a session initialization with a denying access decision $d_{Default}$ that matches the access request pattern from the access request exactly.
    A DEP initializes a local session state by adding the encapsulated access decisions to a local set of access decisions for incoming or outgoing data frames.
    \begin{figure}
        \centering
        \includegraphics[width=1.0\linewidth]{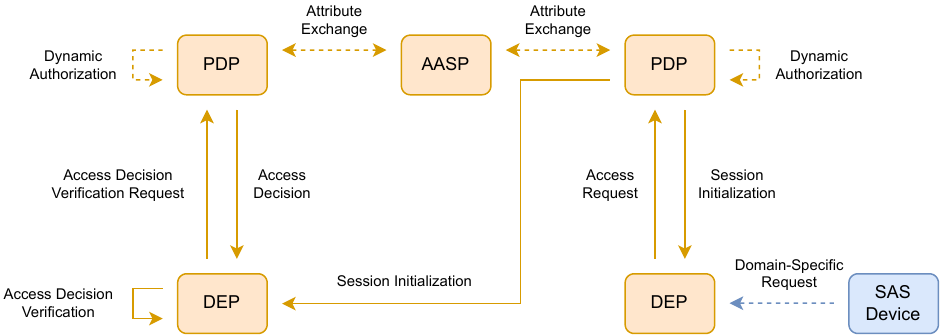}
        \caption{Components and exchanged messages involved in an access request, session initialization, and access decision verification.}
        \label{fig:protocols_enforcement_initialization}
    \end{figure}

    \item[Access Decision Verification] The optional server-aided access decision verification enables a DEP to verify a session initialization.
    This optional step is used to reduce the trust in a single PDP.
    The verification is shown in \autoref{fig:protocols_enforcement_initialization}.
    To initiate the verification, a DEP sends an access verification request to a PDP.
    The request contains the flow pattern of the access decision to be verified.
    The PDP sends the corresponding access decisions to the requesting DEP without initializing sessions at the $nexthop$ DEPs.
    The DEP performs the verification by checking if any conflicts exist between the session initialization and received access decisions.
    
    \item[Payload Exchange] The goal of the payload exchange is to securely exchange domain-specific data frames between SAS devices.
    It is initiated by an SAS device via delivery of a domain-specific data frame to its DEP.
    On receipt of a data frame $f$, the sender's DEP identifies all matching access control decisions $d_1, \dots, d_n$ via flow pattern matching with the access request pattern of $f$.
    As discussed in \autoref{sec:approach:evaluation}, a DEP uses the access decision with the most specific flow pattern match, or a composite decision in case of a conflict.
    If no matching access decision is available, a DEP initiates an access request procedure.
    As soon as the access decision is available, the DEP executes the decision enforcement function $E_{DEC}$.
    If the access decision is granting, the DEP encapsulates the domain-specific data frame in a payload exchange request, and forwards the request to each $DEP_i \in nexthop$.
    
    On receipt of the payload exchange request, the receiving DEP fetches the most specific access decision and evaluates $E_{DEC}$.
    If the access decision is granting and the receiving $DEP \in nexthop$, the domain-specific data frame is extracted from the payload exchange request and forwarded to the receiving SAS device.
    Otherwise, the payload exchange request is discarded.
    A successful payload exchange between two SAS devices is shown in \autoref{fig:protocols_enforcement_payloadexchange}.
    \begin{figure}
        \centering
        \includegraphics[width=1.0\linewidth]{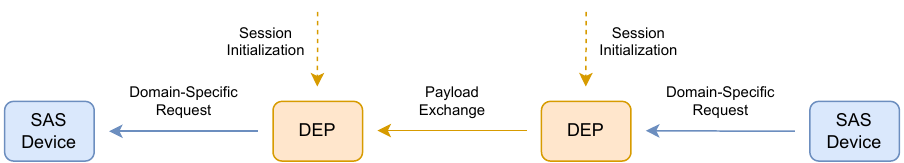}
        \caption{Components and exchanged messages involved in a payload exchange between two SAS devices.}
        \label{fig:protocols_enforcement_payloadexchange}
    \end{figure}
\end{description}

\section{Evaluation}
\label{sec:evaluation}
The evaluation of RTS-ABAC is performed experimentally using industrial SAS devices and a testbed based on off-the-shelf hardware.
The experiments utilize a software implementation of RTS-ABAC.
The software is implemented component-wise using the object-oriented high-level programming languages Java and Kotlin.
The software implementation of RTS-ABAC is published open source on GitHub \cite{gitcasc} under the European Union Public Licence (EUPL) \cite{eupl}.

\subsection{Performance Analysis}
The objective of the performance analysis is to demonstrate that RTS-ABAC is a viable solution to secure time-critical message exchanges.
For this purpose, we conduct an experimental estimation of message exchange latencies using a testbed based on off-the-shelf hardware.

\subsubsection{Experimental Setup}
\label{sec:evaluation:performance:setup}
\begin{figure}
    \centering
    \includegraphics[width=1.0\linewidth]{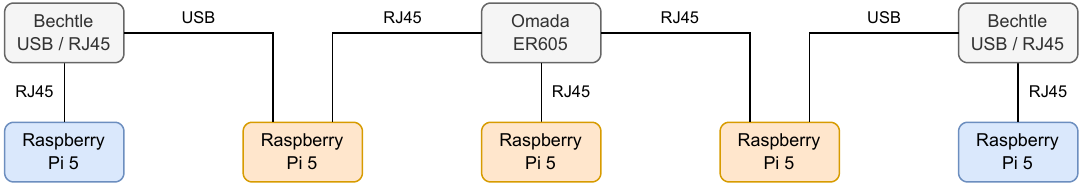}
    \caption{Devices and network topology of the performance analysis testbed.}
    \label{fig:performance_analysis_topology}
\end{figure}

The testbed of the performance analysis consists of eight devices.
The hardware devices and network topology of the testbed is shown in \autoref{fig:performance_analysis_topology}.
Two Raspberry Pi 5 mimic SAS devices, two Raspberry Pi 5 protect the SAS devices as BITW DEPs, and a fifth Raspberry Pi 5 provides the services of the PASP, AASP, and PDP.
We interconnect the devices using Ethernet over twisted-pair.
Since the Raspberry Pi 5 only possesses a single on-board RJ45 Ethernet connector, we additionally use a USB-A to RJ45 Ethernet adapter made by Bechtle for both DEPs.

\subsubsection{Procedure \& Results}
\label{sec:evaluation:performance:procedure}
\begin{figure}
    \centering
    \includegraphics[width=1.0\linewidth]{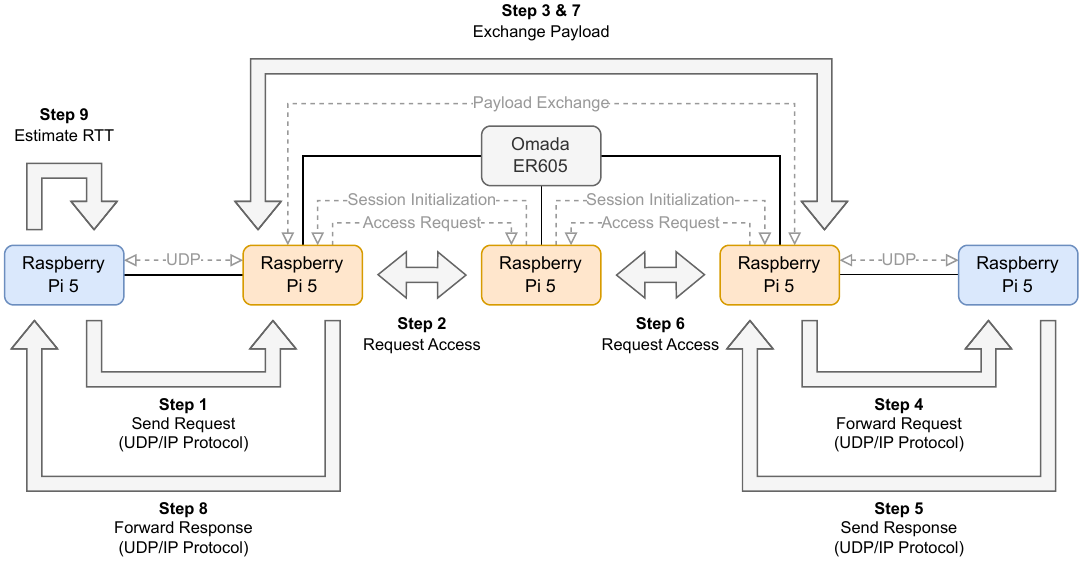}
    \caption{Sequence of events of the experimental message exchange latency estimation.}
    \label{fig:performance_analysis_latency_steps}
\end{figure}

To demonstrate the viability of RTS-ABAC with regard to securing time critical message exchanges, we evaluate the extent to which RTS-ABAC is capable of handling different message types.
For this purpose, we conduct an experiment to estimate the end-to-end communication latency of a message exchange between two domain entities.
The sequence of events and the corresponding messages exchanged are shown in \autoref{fig:performance_analysis_latency_steps}.
We realize the message exchange latency estimation by implementing a benchmark program and deploying it to the domain entities.
The benchmark program is implemented in Python and is published open source on GitHub \cite{gitcasc} alongside the implementation of RTS-ABAC.
The program estimates the end-to-end latency between the two domain entities based on the round-trip time (RTT) of a bidirectional message exchange.
We utilize the user datagram protocol (UDP) as a message exchange protocol for the experiment in order to avoid external latency influences, such as the flow control and congestion control of the transmission control protocol (TCP).
To measure the RTT, the so-called active benchmark entity sends a timestamp to the passive benchmark entity.
The passive entity replies to the message with the received timestamp.
Thus, after receiving the response from the passive entity, the active entity is able to calculate the RTT by subtracting the received timestamp from the current timestamp.
As a consequence, no time synchronization is required between the domain entities.
Furthermore, under the assumption of symmetric transmission times, the accuracy of the RTT measurements only depends on the accuracy of the active entity's system clock.
To avoid an offset caused by the router's buffering and forwarding strategy, messages are sent sequentially, i.e., the active entity waits for the arrival of a response before sending another timestamp message.
Moreover, to compensate for fluctuations in the RTT measurements and to increase the confidence in the RTT estimation, the active entity repeats the RTT measurement procedure.

The results of the latency estimation experiment are shown in \autoref{tab:rtt_metrics} and \autoref{tab:rtt_share}.
The raw data of the individual experiment runs are published open source on GitHub \cite{gitcasc} alongside the implementation of RTS-ABAC.
Since RTS-ABAC is cryptography-dependent but scheme-agnostic, we conduct the latency estimation procedure for different authentication algorithms.
For each of the algorithms, the benchmark program sends 5000 sequential packets to estimate the RTT.
Based on the measurements, we calculate the arithmetic mean, median, standard deviation, and extrema-related values for each algorithm.
Furthermore, we calculate the throughput of the DEPs in packets per second (PPS), and the cumulative share of the 5000 packets in the message types of IEC 61850 listed in \autoref{tab:message_types}.

To measure the latency offset caused by packet capturing and forwarding, we perform the initial latency estimation without authentication, authorization, or access control.
The results indicate that the bidirectional message exchange between two domain entities requires 0.644 ms on average, which leads to a PEP throughput of 1487.2 sequential PPS.
Without authentication, authorization, and access control in place, the measured RTTs are consistent, and each bidirectional message exchange is finished in less than 6 ms RTT.

To measure the influence of the authorization and access control, we perform a latency estimation with authorization and access control but without authentication.
For this purpose, we implement a so-called no-operation (NoOp) authenticator, which processes the packets without signing or verifying them.
The measurements using the NoOp authenticator show that the authorization and access control workflow leads to an RTT increase of 1.555 ms on average.
Moreover, the increased range of minimum and maximum time indicates that authorization and access control lead to an increase in RTT fluctuation.
Nevertheless, 99.9 \% of the messages are exchanged in less than 6 ms RTT.

In order to evaluate the performance of RTS-ABAC in combination with symmetric cryptography, we conduct a latency estimation for HMAC authentication based on SHA-512.
The results show that 99.82 \% of the messages are exchanged in less than 6 ms RTT.
The remaining nine packets or 0.18 \% of the 5000 packets satisfy the 40 ms time constraint of the IEC 61850 message type 1B.
By employing HMAC authentication, RTS-ABAC achieves a throughput of 435.2 PPS at the DEPs.

To evaluate the performance of RTS-ABAC in combination with public key cryptography (PKC), we conduct latency estimations for two PKC algorithms.
For the latency estimations we use the Ed25519 algorithm and the RSA-2048 algorithm.
Both PKC approaches do not satisfy the time constraints of the message types 1A and 4.
However, Ed25519 and RSA-2048 satisfy the time constraint of the message type 1B.
The data indicates that the RTTs of the PKC approaches are subject to fluctuations with higher magnitude compared to the fluctuations caused by symmetric cryptography.
Furthermore, the throughput of the DEPs is reduced by more than 78 \% compared to the HMAC authentication.
\begin{table}
    \centering
    \small
    \caption{Results of the RTT estimation based on 5000 measurements per algorithm.}
    \label{tab:rtt_metrics}
    \begin{tabular}{c c c c c c c c}
    \toprule
    Algorithm & Mean & Median & Deviation & \multicolumn{4}{c}{Extrema}\\
    \cmidrule(lr){5-8}
    & $\bar{x}$ & $\widetilde{x}$ & $\sigma$ & Min & Max & Range & Mid-Range\\
    \midrule
    None                &  0.644 &  0.612 & 0.122 &  0.560 &  4.810 &  4.250 &  2.685 \\
    RTS-ABAC + NoOp     &  2.199 &  2.026 & 0.430 &  1.676 &  9.807 &  8.131 &  5.741 \\
    RTS-ABAC + HMAC     &  2.276 &  2.160 & 0.491 &  1.738 &  9.350 &  7.612 &  5.544 \\
    RTS-ABAC + Ed25519  & 10.478 & 10.721 & 1.465 &  8.426 & 15.775 &  7.348 & 12.101 \\
    RTS-ABAC + RSA-2048 & 12.221 & 12.063 & 1.591 & 10.453 & 57.004 & 46.551 & 33.729 \\
    \bottomrule
    \end{tabular}
\end{table}
\begin{table}
    \centering
    \small
    \caption{Throughput and cumulative message type share of the analyzed algorithms.}
    \label{tab:rtt_share}
    \begin{tabular}{c c c c c c}
    \toprule
    Algorithm & Throughput & \multicolumn{4}{c}{Cumulative Share in Message Types}\\
    \cmidrule(lr){3-6}
     & & 1A / 4 & 1B & 2 & 3 / 5 / 6\\
     & & $\leq 6~ms$ & $\leq 40~ms$ & $\leq 200~ms$ & $\leq 1000~ms$\\
    \midrule
    None                & 1487.2 PPS & 5000 (100   \%) & 5000 (100 \%) & 5000 (100 \%) & 5000 (100 \%) \\
    RTS-ABAC + NoOp     &  448.8 PPS & 4995 (99.9 \%) & 5000 (100 \%) & 5000 (100 \%) & 5000 (100 \%) \\
    RTS-ABAC + HMAC     &  435.2 PPS & 4991 (99.82 \%) & 5000 (100 \%) & 5000 (100 \%) & 5000 (100 \%) \\
    RTS-ABAC + Ed25519  &   94.9 PPS &    0 (0 \%) & 5000 (100 \%) & 5000 (100 \%) & 5000 (100 \%) \\
    RTS-ABAC + RSA-2048 &   81.4 PPS &    0 (0 \%) & 4998 (99.96 \%) & 5000 (100 \%) & 5000 (100 \%) \\
    \bottomrule
    \end{tabular}
\end{table}

\subsection{Demonstration of Applicability}
We conduct a demonstration of applicability with industrial SAS devices in our smart grid cybersecurity laboratory \cite{Elbez2025}.
The demonstration serves to complement the performance analysis by demonstrating that RTS-ABAC is a viable solution for enhancing the communication security in a newly constructed or retrofitted SAS.
Accordingly, we demonstrate that the SAS behavior and functionality are not negatively influenced by RTS-ABAC, and that SAS devices protected by RTS-ABAC are capable of providing their services and exchanging information.

\subsubsection{Experimental Setup}
\label{sec:evaluation:compatibility:setup}
The hardware devices used for the experiment are listed in \autoref{tab:lab_hardware}.
Nine of these devices are industrial SAS devices, which are organized in three so-called bays.
Each bay consists of an MU, a protection IED, and an I/O box.
We conduct the experiment for each of the three bays individually.
To simulate a power grid, we use a relay test device from Omicron to generate three-phase electric power, which is measured by the MUs.
Each I/O box is connected to an Omicron process simulator, which mimics the behavior of a substation circuit breaker.
Besides these SAS-related devices, Raspberry Pi 5 computers provide the services of the RTS-ABAC components.
\begin{table}
    \centering
    \small
    \caption{Hardware used for the laboratory-based experimental demonstration of applicability.}
    \label{tab:lab_hardware}
    \begin{tabular}{c | c c c}
    \toprule
    & Manufacturer & Device & Task\\
    \midrule
    \multirow{3}{*}{Bay I} & General Electric & Reason MU320   & Process Bus MU   \\
                           & General Electric & Multilin F60   & Protection IED   \\
                           & Siemens          & SIPROTEC 6MD84 & Input/Output Box \\
    \cmidrule(lr){1-4}
    \multirow{3}{*}{Bay II} & SEL         & SEL-401        & Process Bus MU   \\
                            & Hitachi ABB & REL670         & Protection IED   \\
                            & Siemens     & SIPROTEC 6MD84 & Input/Output Box \\
    \cmidrule(lr){1-4}
    \multirow{3}{*}{Bay III} & Siemens & SIPROTEC 6MU85 & Process Bus MU   \\
                             & Siemens & SIPROTEC 7SX85 & Protection IED   \\
                             & Siemens & SIPROTEC 6MD84 & Input/Output Box \\
    \cmidrule(lr){1-4}
    \multirow{2}{*}{Grid Simulation} & OMICRON & CMC 356 & Universal Relay Test Set  \\
                                     & OMICRON & Process Simulator & Circuit Breaker \\
    \cmidrule(lr){1-4}
    \multirow{3}{*}{Networking} & Weidmüller & IE-SW-SL28M-SV   & Industrial Ethernet Switch \\
                                & FS.COM     & UMC-1F1T         & Ethernet Media Converter   \\
                                & Bechtle    & ARTICONA Adapter & USB-A to RJ45 Adapter      \\
    \cmidrule(lr){1-4}
    RTS-ABAC & Raspberry Pi Ltd & Raspberry Pi 5 8GB & DEP, PDP, PASP, \& AASP \\
    \bottomrule
    \end{tabular}
\end{table}

\begin{figure}
    \centering
    \includegraphics[width=1.0\linewidth]{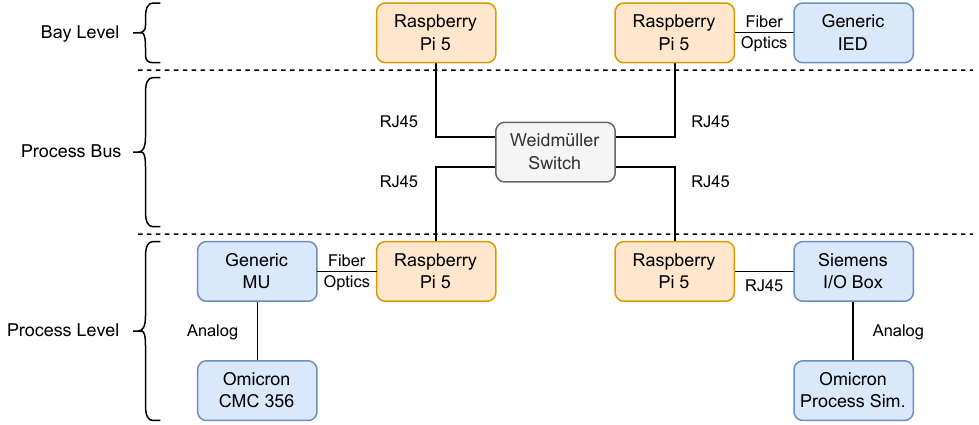}
    \caption{Network topology of the laboratory-based demonstration of applicability.}
    \label{fig:lab_topology}
\end{figure}

The network topology of the devices is shown in \autoref{fig:lab_topology}.
To improve the readability of the shown topology, the USB-A to RJ45 Ethernet adapters and Ethernet media converters are omitted from the figure.
In accordance with the layered SAS architecture shown in \autoref{fig:substation_architecture}, we introduce a layering of devices for the setup of the laboratory experiment.

\subsubsection{Procedure \& Results}
\label{sec:evaluation:compatibility:procedure}
\begin{figure}
    \centering
    \includegraphics[width=1.0\linewidth]{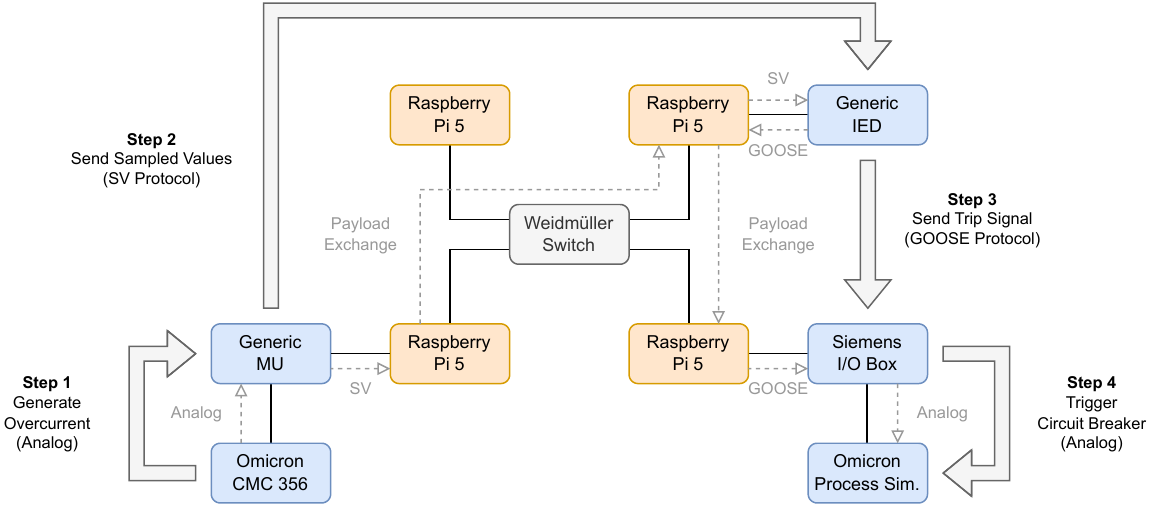}
    \caption{Sequence of events of the laboratory-based demonstration of applicability.}
    \label{fig:lab_steps}
\end{figure}

At the initial state of the experiment, the voltages and currents of all three phases generated by the Omicron CMC 356 are within certain boundaries to be detected as a valid grid situation.
Accordingly, the circuit breakers connected to the I/O boxes are closed and allow electric power to flow.
The procedure of the experiment comprised four key events.
The sequence of events and the corresponding messages exchanged between the devices are shown in \autoref{fig:lab_steps} and are discussed in the following:
\begin{description}
    \item[Step 1 "Generate Overcurrent":]
    To start the experiment, we adjust the generated three-phase electric power of the relay test device.
    We set the current to a level that represents an overcurrent situation in our laboratory grid.
    This situation is communicated to the MUs via a direct analog connection, i.e., the MUs measure the voltages and currents of the three phases.

    \item[Step 2 "Send Sampled Values":]
    The MUs sample the voltage and current values provided by our relay test device.
    The MUs send the sampled values to the protection IEDs using the SV protocol.
    As the MUs are protected by a DEP, the outgoing SV frames are captured by the DEPs and processed as discussed in \autoref{sec:approach:protocol_enforcement}.
    The authenticated and authorized payload exchange messages, which contain the SV frames, are then forwarded to the DEPs of the IEDs.

    \item[Step 3 "Send Trip Signal":]
    The IEDs receive and process the SV frames, and detect the overcurrent situation.
    To resolve the overcurrent situation, the IEDs send GOOSE frames to their I/O boxes to open the corresponding circuit breakers.
    As the IEDs are protected by a DEP, the outgoing GOOSE frames are captured by the DEPs and processed as discussed in \autoref{sec:approach:protocol_enforcement}.
    The authenticated and authorized payload exchange messages, which contain the GOOSE frames, are then forwarded to the DEPs of the I/O boxes.

    \item[Step 4 "Trigger Circuit Breaker":]
    The I/O boxes receive the GOOSE frames and, as the GOOSE frames signal an overcurrent situation, the I/O boxes use an analog signal to open their circuit breakers.
    At the end of the experiment, we are able to verify electrically and visually that all circuit breakers are open.
\end{description}

\subsection{Discussion \& Limitations}
As the chain of events successfully propagates through our laboratory SAS network, we demonstrate that RTS-ABAC does not disrupt the operation of the MUs, IEDs, and I/O boxes.
Since the security-related components are deployed to inexpensive off-the-shelf hardware, we are able to demonstrate that RTS-ABAC is a feasible solution for SAS environments, not only security-wise and performance-wise, but also cost-wise.
Due to its BITW concept, no adaptations have to be made to the SAS devices.
Thus, the interoperability and interchangeability requirements of an IEC 61850 substation remain satisfied, while the communication security is enhanced.
This indicates that RTS-ABAC is a viable solution for the retrofitting of existing substations.

The performance analysis reveals that the attribute-based authorization and access control workflow of RTS-ABAC has only a limited impact on the overall message exchange latency.
Accordingly, the results emphasize the appropriateness of expressive and flexible yet computationally expensive access control approaches, such as ABAC, even in time and resource constrained environments.
Nevertheless, the results of the performance analysis indicate that computational performance is the primary challenge for the deployment of cybersecurity approaches in an SAS.
The findings show that lightweight yet inflexible symmetric cryptography seems to be the only viable solution to authenticate low latency message exchanges at the moment.
We anticipate that the required computation time can be decreased, leading to an increase in message throughput, through the utilization of hardware accelerators for cryptographic algorithms.
However, factors such as algorithm compatibility, costs per acceleration unit, and computation time consistency may result in a less beneficial influence on the system than currently expected.

The evaluation demonstrates that RTS-ABAC is capable of securing application protocols of an SAS, as well as multipurpose transport protocols.
However, by employing RTS-ABAC in a substation the attack surface of the corresponding SAS might change.
While IEDs, MUs, and other SAS devices benefit from being protected by using our BITW approach, the increased total number of devices in the SAS and their communication relations lead to an increased risk of denial-of-service (DoS) attacks.
In particular, the dependence on certain locally or globally centralized components, such as PDP, PASP, and AASP, might introduce new attack vectors for SAS adversaries.

Furthermore, network time protocols and lower-layer network management protocols, such as the address resolution protocol (ARP), parallel redundancy protocol (PRP), and media redundancy protocol (MRP), are not secured but forwarded by the DEPs.
The protocols are bypassed since they provide services not only to SAS devices but also to auxiliary intermediate devices, including network switches and routers, and as the operation of these protocols is susceptible to temporal inconsistencies resulting from authentication, authorization, and access control.
Nevertheless, the usage of static bypass rules in the DEPs suggests that incompatible or legacy devices and protocols could continue their operation in retrofitted substations even if RTS-ABAC is deployed.

\section{Conclusions \& Outlook}
\label{sec:conclusion}
To address the increasing relevance of cybersecurity for smart grid systems and to overcome the limitations of existing standards like IEC 61850 and IEC 62351, we present RTS-ABAC, a novel real-time server-aided attribute-based authorization and access control approach.
RTS-ABAC enables the creation, management, storage, distribution, evaluation, and enforcement of ABAC policies for time-critical and time-variable distributed systems.
We introduce the concept of time-dependency for ABAC attributes and policies.
Furthermore, we discuss methods to manage, distribute, and enforce such expressive and flexible yet computationally expensive access control policies and access control decisions.
We propose a delegated attribute-based authorization protocol, which is responsible for the policy creation, management, storage, distribution, and evaluation of RTS-ABAC policies.
Finally, we additionally introduce a delegated decision enforcement protocol, which is responsible for the exchange, verification, and enforcement of access control decisions.

RTS-ABAC is implemented in software and is published open source on GitHub \cite{gitcasc} under the European Union Public Licence (EUPL) \cite{eupl}.
Based on the implementation we assess the applicability of RTS-ABAC for SAS communication protocols by conducting a performance analysis.
The results of the performance analysis are published open source alongside the software implementation \cite{gitcasc}.
The performance analysis demonstrates the ability of RTS-ABAC to secure time-critical message exchanges, as up to 99.82 \% of authenticated, authorized, and access controlled network packets achieve an RTT below 6 ms.
In particular, the performance analysis shows that the time required to perform cryptographic authentication procedures exceeds the time to evaluate and enforce RTS-ABAC policies.
Furthermore, the performance analysis identifies the advantages and disadvantages of different authentication schemes with regard to satisfied time constraints.
In accordance with the related literature, we identify the strict time constraints of low latency communication in an SAS as a key challenge for cybersecurity.

We demonstrate the applicability of RTS-ABAC for SAS devices in our smart grid cybersecurity laboratory \cite{Elbez2025}.
The laboratory-based demonstration of applicability shows the ability of RTS-ABAC to secure the GOOSE and SV protocol between industrial-grade IEDs, MUs, and I/O boxes made by ABB, General Electric, Siemens, and SEL.
Furthermore, we show that RTS-ABAC is a feasible solution for SAS environments, not only security-wise and performance-wise, but also cost-wise and due to its highly-compatible BITW concept, which allows retrofitting of existing systems.
Accordingly, the results of the evaluation indicate that RTS-ABAC is a viable approach to enhance the communication security in a newly constructed or retrofitted substation.

While SAS-typical cyberattacks can be mitigated by employing RTS-ABAC, we also discuss the change of the attack surface, leading to an increased risk of DoS as a result of the additional components and protocols deployed in an SAS.
To simplify the architectural complexity of RTS-ABAC, reduce its attack surface, and reduce the overall costs of deployment, we propose the integration of RTS-ABAC into network switches as an alternative realization.
For this purpose, further research could investigate the potential benefits of realizing RTS-ABAC through the use of software-defined networking (SDN) solutions.
SDN-based realization could aggregate the tasks of multiple DEPs by deploying a virtual DEP for each port of a network switch.
Furthermore, distributed SDN controllers might provide the PASP, AASP, and PDP services.
While the proposed PASP entities provide policy management services for human operators, future research could investigate how RTS-ABAC might benefit from the utilization of artificial intelligence (AI).
The integration of AI-based intrusion detection could facilitate the creation and modification of security policies that are enforced within an SAS, thereby enabling the mitigation of a wider range of cyberattacks in a timelier manner.
In addition to the deployment in an SAS, further research is required to evaluate the applicability of RTS-ABAC for other time-critical distributed systems.
Therefore, we propose its evaluation in systems that have similar requirements as an SAS.
Systems that potentially benefit from the enhanced communication security provided by RTS-ABAC include industry 4.0, robotics, avionics, and medical systems.

\section*{Acknowledgments}
This work was supported by funding from the topic Engineering Secure Systems of the Helmholtz Association (HGF) and by KASTEL Security Research Labs (structure 46.23.02).
\printbibliography

@TechReport{Hu2014,
  author      = {Hu, Vincent C. and Ferraiolo, David and Kuhn, Rick and Schnitzer, Adam and Sandlin, Kenneth and Miller, Robert and Scarfone, Karen},
  institution = {National Institute of Standards and Technology},
  title       = {Guide to Attribute Based Access Control (ABAC) Definition and Considerations},
  year        = {2014},
  month       = jan,
  number      = {NIST Special Publication 800-162},
  doi         = {10.6028/nist.sp.800-162},
}

@TechReport{JTF2020,
  author      = {Task Force Interagency Working Group, Joint},
  institution = {National Institute of Standards and Technology},
  title       = {Security and Privacy Controls for Information Systems and Organizations},
  year        = {2020},
  month       = sep,
  number      = {NIST Special Publication 800-53,Rev.5},
  doi         = {10.6028/nist.sp.800-53r5},
}

@TechReport{Stouffer2023,
  author      = {Stouffer, Keith and Pease, Michael and Tang, CheeYee and Zimmerman, Timothy and Pillitteri, Victoria and Lightman, Suzanne and Hahn, Adam and Saravia, Stephanie and Sherule, Aslam and Thompson, Michael},
  institution = {National Institute of Standards and Technology},
  title       = {Guide to Operational Technology (OT) Security},
  year        = {2023},
  number      = {NIST Special Publication 800-82,Rev.3},
  doi         = {10.6028/nist.sp.800-82r3},
}

@InProceedings{Ishchenko2018,
  author    = {Ishchenko, Dmitry and Nuqui, Reynaldo},
  booktitle = {2018 IEEE/PES Transmission and Distribution Conference and Exposition (T\&D)},
  title     = {Secure Communication of Intelligent Electronic Devices in Digital Substations},
  year      = {2018},
  month     = apr,
  publisher = {IEEE},
  doi       = {10.1109/tdc.2018.8440438},
}

@Article{Rodriguez2021,
  author    = {Rodriguez, Mikel and Lazaro, Jesus and Bidarte, Unai and Jimenez, Jaime and Astarloa, Armando},
  journal   = {IEEE Access},
  title     = {A Fixed-Latency Architecture to Secure GOOSE and Sampled Value Messages in Substation Systems},
  year      = {2021},
  issn      = {2169-3536},
  pages     = {51646--51658},
  volume    = {9},
  doi       = {10.1109/access.2021.3069088},
  publisher = {Institute of Electrical and Electronics Engineers (IEEE)},
}

@InProceedings{Ruland2018,
  author    = {Ruland, Christoph and Sassmannshausen, Jochen},
  booktitle = {2018 IEEE International Conference on Smart Energy Grid Engineering (SEGE)},
  title     = {Firewall for Attribute-Based Access Control in Smart Grids},
  year      = {2018},
  month     = aug,
  publisher = {IEEE},
  doi       = {10.1109/sege.2018.8499306},
}

@Online{oshaSubstation,
  author  = {{Occupational Safety and Health Administration (OSHA)}},
  title   = {Illustrated Glossary - Substations},
  url     = {https://www.osha.gov/etools/electric-power/illustrated-glossary/sub-station},
  lastaccessed = {January 23, 2026},
}

@InProceedings{Burmester2013,
  author    = {Burmester, Mike and Magkos, Emmanouil and Chrissikopoulos, Vassilis},
  booktitle = {2013 IEEE Symposium on Computers and Communications (ISCC)},
  title     = {T-ABAC: An attribute-based access control model for real-time availability in highly dynamic systems},
  year      = {2013},
  month     = jul,
  publisher = {IEEE},
  doi       = {10.1109/iscc.2013.6754936},
}

@InProceedings{Ma2006,
  author    = {Ma, Mingchao and Woodhead, Steve},
  booktitle = {The Sixth IEEE International Conference on Computer and Information Technology (CIT’06)},
  title     = {Constraint-Enabled Distributed RBAC for Subscription-Based Remote Network Services},
  year      = {2006},
  publisher = {IEEE},
  doi       = {10.1109/cit.2006.63},
}

@InProceedings{Elbez2019,
  author    = {Elbez, Ghada and Keller, Hubert B. and Hagenmeyer, Veit},
  booktitle = {Electronic Workshops in Computing},
  title     = {Authentication of GOOSE Messages under Timing Constraints in IEC 61850 Substations},
  year      = {2019},
  month     = sep,
  publisher = {BCS Learning \& Development},
  doi       = {10.14236/ewic/icscsr19.17},
  issn      = {1477-9358},
}

@Article{Ma2006a,
  author    = {Ma, Mingchao and Woodhead, Steve},
  journal   = {Computers \& Security},
  title     = {Authentication delegation for subscription-based remote network services},
  year      = {2006},
  issn      = {0167-4048},
  month     = jul,
  number    = {5},
  pages     = {371--378},
  volume    = {25},
  doi       = {10.1016/j.cose.2006.03.006},
  publisher = {Elsevier BV},
}

@Article{IEC61850P5,
  author  = {{International Electrotechnical Commission}},
  journal = {Communication networks and systems for power utility automation (IEC 61850)},
  title   = {Part 5: Communication requirements for functions and device models},
  year    = {2014},
}

@Article{IEC61850P6,
  author  = {{International Electrotechnical Commission}},
  journal = {Communication networks and systems for power utility automation (IEC 61850)},
  title   = {Part 6: Configuration description language for communication in power utility automation systems related to IEDs},
  year    = {2020},
}

@Article{IEC61850P8,
  author  = {{International Electrotechnical Commission}},
  journal = {Communication networks and systems for power utility automation (IEC 61850)},
  title   = {Part 8-1: Specific communication service mapping (SCSM) - Mappings to MMS (ISO 9506-1 and ISO 9506-2) and to ISO/IEC 8802-3},
  year    = {2022},
}

@Article{IEC62351P8,
  author  = {{International Electrotechnical Commission}},
  journal = {Power systems management and associated information exchange - Data and communications security (IEC 62351)},
  title   = {Part 8: Role-based access control for power system management},
  year    = {2020},
}

@Article{IEC62351P6,
  author  = {{International Electrotechnical Commission}},
  journal = {Power systems management and associated information exchange - Data and communications security (IEC 62351)},
  title   = {Part 6: Security for IEC 61850},
  year    = {2020},
}

@Book{Padilla2015,
  author    = {Padilla, Evelio},
  publisher = {Wiley},
  title     = {Substation Automation Systems: Design and Implementation},
  year      = {2015},
  isbn      = {9781118987216},
  month     = oct,
  doi       = {10.1002/9781118987216},
}

@Online{Oasis2013,
  author  = {{OASIS Open}},
  title   = {eXtensible Access Control Markup Language (XACML) Version 3.0},
  url     = {https://docs.oasis-open.org/xacml/3.0/xacml-3.0-core-spec-os-en.pdf},
  lastaccessed = {January 23, 2026},
  year    = {2013},
}

@Online{canada2021,
  author  = {{Communications Security Establishment Canada}},
  month   = dec,
  title   = {Cyber threat bulletin: Cyber threat to operational technology},
  url     = {https://open.canada.ca/data/dataset/98bad300-28f1-49b9-9b34-2d46de4c9a58},
  lastaccessed = {January 23, 2026},
  year    = {2021},
}

@Online{bbc2010,
  author  = {Jonathan Fildes},
  title   = {Stuxnet worm targeted high-value Iranian assets},
  url     = {https://www.bbc.com/news/technology-11388018},
  lastaccessed = {January 23, 2026},
  year    = {2010},
}

@Online{reuters2012,
  author  = {Jim Finkle},
  title   = {Insiders suspected in Saudi cyber attack},
  url     = {https://www.reuters.com/article/technology/exclusive-insiders-suspected-in-saudi-cyber-attack-idUSBRE8860CR},
  lastaccessed = {January 23, 2026},
  year    = {2012},
}

@Online{cisa2021a,
  author  = {{Cybersecurity \& Infrastructure Security Agency (CISA)}},
  title   = {Cyber-Attack Against Uk\-rai\-nian Critical Infrastructure},
  url     = {https://www.cisa.gov/news-events/ics-alerts/ir-alert-h-16-056-01},
  lastaccessed = {January 23, 2026},
  year    = {2021},
}

@Online{reuters2016,
  author  = {Natalia Zinets},
  title   = {Ukraine hit by 6,500 hack attacks, sees Russian cyberwar},
  url     = {https://www.reuters.com/article/us-ukraine-crisis-cyber-idUSKBN14I1QC},
  lastaccessed = {January 23, 2026},
  year    = {2016},
}

@Online{cisa2021b,
  author  = {{Cybersecurity \& Infrastructure Security Agency (CISA)}},
  title   = {CrashOverride Malware},
  url     = {https://www.cisa.gov/news-events/alerts/2017/06/12/crashoverride-malware},
  lastaccessed = {January 23, 2026},
  year    = {2021},
}

@Online{johnson2017,
  author  = {Johnson, Blake and Caban, Dan and Krotofil, Marina and Scali, Dan and Brubaker, Nathan and Glyer, Christopher},
  title   = {Attackers Deploy New ICS Attack Framework "TRITON" and Cause Operational Disruption to Critical Infrastructure},
  url     = {https://cloud.google.com/blog/topics/threat-intelligence/attackers-deploy-new-ics-attack-framework-triton},
  lastaccessed = {January 23, 2026},
  year    = {2017},
}

@Article{Fang2012,
  author    = {Fang, Xi and Misra, Satyajayant and Xue, Guoliang and Yang, Dejun},
  journal   = {IEEE Communications Surveys \& Tutorials},
  title     = {Smart Grid — The New and Improved Power Grid: A Survey},
  year      = {2012},
  issn      = {1553-877X},
  number    = {4},
  pages     = {944--980},
  volume    = {14},
  doi       = {10.1109/surv.2011.101911.00087},
  publisher = {Institute of Electrical and Electronics Engineers (IEEE)},
}

@Online{eupl,
  author  = {{European Union}},
  title   = {{European Union Public Licence (EUPL) Version 1.2}},
  url     = {https://joinup.ec.europa.eu/collection/eupl},
  lastaccessed = {January 23, 2026},
  year    = {2017},
}

@Online{gitcasc,
  author  = {Moritz Gstuer},
  title   = {{Real-Time Server-Aided Attribute-Based Authorization \& Access Control for Substation Automation Systems}},
  url     = {https://github.com/gstuer/RTS-ABAC},
  lastaccessed = {January 23, 2026},
  year    = {2026},
}

@Article{Kim2022,
  author    = {Kim, Hwimin and Kim, Dae-Kyoo and Alaerjan, Alaa},
  journal   = {IEEE Transactions on Dependable and Secure Computing},
  title     = {ABAC-Based Security Model for DDS},
  year      = {2022},
  issn      = {2160-9209},
  month     = sep,
  number    = {5},
  pages     = {3113--3124},
  volume    = {19},
  doi       = {10.1109/tdsc.2021.3085475},
  publisher = {Institute of Electrical and Electronics Engineers (IEEE)},
}

@MastersThesis{Gstuer2025,
  author    = {Gstür, Moritz},
  school    = {Karlsruher Institut für Technologie (KIT)},
  title     = {Certificateless Attribute-Based Server-Aided Cryptosystem for Substation Automation Systems (CASC-SAS)},
  year      = {2025},
  doi       = {10.5445/IR/1000182038},
  language  = {english},
  publisher = {{Karlsruher Institut für Technologie (KIT)}},
}

@InProceedings{Elbez2025,
  author     = {Elbez, Ghada and Sánchez, Gustavo and Canbolat, Sine and Corallo, Sophie and Fruböse, Clemens and Lanzinger, Florian and Kellerer, Nicolai and Keppler, Gustav and Neumeister, Felix and Beckert, Bernhard and Koziolek, Anne and Zitterbart, Martina and Hagenmeyer, Veit},
  booktitle  = {Proceedings of the 16th ACM International Conference on Future and Sustainable Energy Systems},
  title      = {Insights and Lessons Learned from a Realistic Smart Grid Testbed for Cybersecurity Research},
  year       = {2025},
  month      = jun,
  pages      = {805--812},
  publisher  = {ACM},
  series     = {E-Energy ’25},
  collection = {E-Energy ’25},
  doi        = {10.1145/3679240.3734649},
}
\end{document}